\documentclass[aps,pre,twocolumn,showpacs,amsfonts,amsmath,amssymb,superscriptaddress]{revtex4-1}

\usepackage{hyperref}
\usepackage{graphicx}			
\usepackage{color}
\usepackage{fancyvrb}
\usepackage{tikz}
\usepackage[normalem]{ulem}

\graphicspath{{Figures/},.}

\newcommand{\ie}{\textit{i.e.}}
\newcommand{\eg}{\textit{e.g.}}
\newcommand{\etal}{\textit{et~al.}}
\newcommand*\mycirc[1]{\textcircled{#1}}
  
\definecolor{black}{rgb}{0,0,0}
\definecolor{skyblue}{rgb}{0.337255,0.70588,0.91373}
\definecolor{vermilion}{rgb}{0.83529,0.3863,0}
\definecolor{gray}{rgb}{0.8,0.8,0.8}

\begin{document}
\title{Fair sharing of resources in a supply network with constraints}

\author{Rui Carvalho}
\email{rui@maths.qmul.ac.uk}
\affiliation{School of Mathematical Sciences, Queen Mary University of London, Mile End Road, London E1 4NS, U.K.}

\author{Lubos Buzna}
\email{buzna@frdsa.fri.uniza.sk}
\affiliation{University of Zilina, Univerzitna 8215/5, 01026 Zilina, Slovakia}

\author{Wolfram Just}
\affiliation{School of Mathematical Sciences, Queen Mary University of London, Mile End Road, London E1 4NS, U.K.}

\author{Dirk Helbing}
\affiliation{ETH Zurich, CLU E1, Clausiusstr. 50, 8092 Zurich, Switzerland}
\affiliation{Santa Fe Institute, 1399 Hyde Park Road, Santa Fe, NM 87501, USA}
\affiliation{Collegium Budapest - Institute for Advanced Study, Szenth\'{a}roms\'{a}g u. 2, 1014 Budapest, Hungary}

\author{David K. Arrowsmith}
\affiliation{School of Mathematical Sciences, Queen Mary University of London, Mile End Road, London E1 4NS, U.K.}

\begin{abstract}
This paper investigates the effect of network topology on the fair allocation of network resources among a set of agents, an all-important issue for the efficiency of transportation networks all around us. We analyse a generic mechanism that distributes network capacity fairly among existing flow demands. The problem can be solved by semi-analytical methods on a nearest neighbour graph with one source and sink pair, when transport occurs over shortest paths. For this setup, we uncover a broad range of patterns of intersecting shortest paths as a function of the distance between the source and the sink. When the number of intersections is the maximum and the distance between the source and the sink is large, we find  that a fair allocation implies a decrease of at least $50\%$ from the maximum throughput. We also find that the histogram of the flow allocations assigned to the agents decays as a power-law with exponent -1. Our semi-analytical framework suggests possible explanations for the well-known reduction of the throughput in fair allocations. It also suggests that the combination of network topology and routing rules can lead to highly uneven (but fair) distributions of resources, a remark of caution to network designers.
\end{abstract}
\maketitle 
\section{Introduction}
\label{sec:intro}
Transportation networks carry the flow of some commodity or entity using routing rules that aim to be efficient and avoid congestion~\cite{Whittle07,Barrat08}. Recently, researchers have focused on the design of optimal transportation networks~\cite{Bohn07,Corson10,Li10,Xia10}, the characterisation of robustness and damage in transport networks~\cite{Carvalho09,Katifori10,Tero10}, and routing strategies that minimize congestion~\cite{Ohira98,Danila06,Sreenivasan07,Marsili09, Danila09}. Despite these efforts, little is known about the effect of network topology and various routing protocols on the competing interests of network users. Indeed, transport frequently involves a very large number of agents, and the operation of the network has to be managed in a smooth and fair way. Moreover, a socially efficient solution, \ie, one that maximizes the sum of the utilities of individual agents, can be difficult to implement because it may be perceived as unfair to some of the agents~\cite{Helbing02,Helbing05,Youn08,Bertsimas11}. 

The fair division of network capacity among agents is a problem related to the procedure of dividing a cake fairly.
Cake cutting, or the fair division of a divisible good among a set of agents, can be analysed using several mathematical interpretations of `fairness'~\cite{Brams96,Robertson98,Barbanel05,Taylor08}. When all $N$ agents agree on the value of each portion, the problem becomes trivial and each agent receives a piece of size $1/N$. If the transport routes intersect along one edge only, the problem of dividing the edge capacity is akin to cutting a cake. However, in most situations of interest, distinct agents share transport routes in sophisticated ways, and the problem is then how to allocate fairly the available capacity on the intricate web of congested edges. 
Two prominent strategies to address this conundrum are the \textit{Max-Min Fairness}~\cite{Bertsekas92}  and the \textit{Proportional Fairness}~\cite{Kelly98} allocations, and this paper focuses on the former. A set of  flows is Max-Min Fair if no flow can be increased without simultaneously decreasing another flow that is already less than or equal to the former. To oversimplify, an allocation is Max-Min Fair if the wealthy can only get wealthier by making the poor even poorer. 

The related problem of network congestion control aims at devising mechanisms for the allocation of resources in transport systems~\cite{Kelly98,Srikant03}. There is a long line of work on addressing the challenge of congestion control with the Max-Min Fair allocation, mainly by researchers working on communication networks.  Indeed, the fair allocation of capacity has been extensively studied in the context of flow control in IP packet switching networks, such as the Internet and ATM networks~\cite{Bertsekas92,Kelly98,Kelly00,Kleinberg01,Massoulie02,Srikant03}. Moreover, researchers have taken a wide range of approaches to the allocation of communication rates, encompassing both non pre-allocated~\cite{Megiddo74,Megiddo77,Saad07} and fixed~\cite{Jaffe81, Bertsekas92} paths. 
Despite these efforts, to the best of our knowledge, previous approaches have focused on the design of algorithms ~\cite{Megiddo74,Megiddo77,Jaffe81,Bertsekas92,Kelly98,Kleinberg01,Massoulie02,Kumar06,Anshelevich08}, or the computation of upper bounds of network throughput~\cite{Bertsimas11}. In contrast, our results offer a proof of concept that flow control can be analysed semi-analytically in large networks, at least regular ones, and reveal insights into the interplay of network topology, capacity, route selection and distance between a source and a sink, and how these elements in turn affect congestion in fair networks. 

The choices of network topology and routing place obvious constraints on the fair allocation of network resources. If the routing is such that paths do not intersect, each path is assigned the minimum capacity among all its edges and the throughput is the max-flow. In general, however, the choice of routing has an impact on the number of path intersections and thus on the reduction of network throughput in fair allocations. One possible choice of routing is to allow transport over all possible s-t paths. In such case, the Max-Min Fairness algorithm is typically solved by an iterative LP model~\cite{Megiddo77,Nace06}, and the fair allocation to one source-sink pair gives the same throughput as max-flow\cite{Megiddo77}. However, the model is computationally heavy when the number of s-t pairs is large~\cite{Nace08a}. Another possibility, followed in the vast majority of work on Max-Min Fairness~\cite{Bertsekas92,Srikant03,Kleinberg01,Jaffe81,Radunovic07,Nace08} and in this paper, is to fix the paths off-line before the algorithm is run. Fixing the paths is justified, although  network managers are confronted with dynamic real-time traffic conditions, because it is frequently impractical to update the routing schemes according to changes in the distribution of traffic~\cite{Nace06,Nace08}. Hence, network operators often design the best possible static network conditions for a set of estimated average or worst-case demands~\cite{Nace08}. Here we choose to analyse routing over shortest paths, because this is a highly schematic way to model several transport processes in the real world, which is frequently used in the literature~\cite{Costa11,Boccaletti_PR_06}. Some real-world examples motivating its usage are packet transfer on the Internet~\cite{Menezes04}, spatial distribution networks such as sewer systems~\cite{Gastner06a}, gas pipeline networks~\cite{Carvalho09}, commuter rail~\cite{Gastner06a} and subway networks~\cite{Latora01}.

\section{Fair division on networks}
\label{sec:fair_division}

\subsection{How to share network capacity fairly in supply networks}
\label{subsec:water_filling}
\begin{figure}[phtb]
\includegraphics[width=0.48\textwidth]{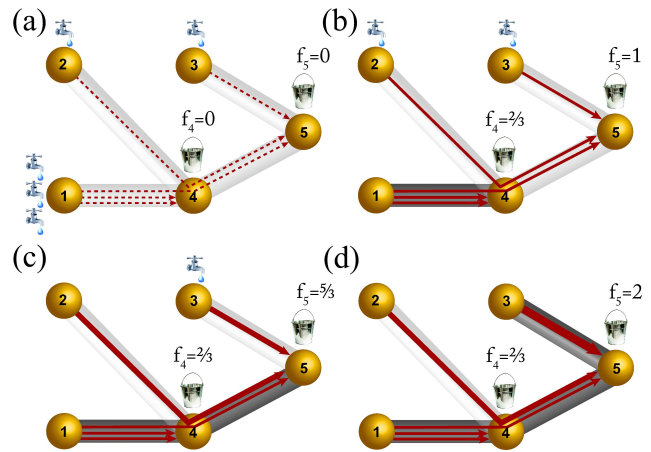}
\caption{\label{fig:Panels}(Colour online). Illustration of the Max-Min Fair (MMF) allocation on a small water pipeline network with three source nodes (identified by \textit{water taps}), two sink nodes (identified by \textit{buckets}), and pipelines with unit capacity. (a) Shows the initial configuration, before the application of the MMF algorithm. Five paths, connect the sources (taps) to the sinks (buckets). (b) At iteration one, the flow is increased uniformly on all taps, until the capacity is reached on pipeline $e_{1,4}$. The path flows that cross $e_{1,4}$ are frozen, the corresponding taps can no longer add to network flow and are thus removed from the figure, and we say that the pipeline is saturated (saturated pipelines are marked in dark gray colour). (c) and (d) The process is repeated for the unsaturated pipelines, until  pipelines $e_{4,5}$ and $e_{3,5}$ are saturated at iterations two and three, respectively. The sink inflow $f_4$ and $f_5$ is displayed on each panel. The final allocation on panel (d) is Max-Min Fair, because no path flow can then be increased without decreasing another path flow already smaller.}
\end{figure}

The Max-Min Fair (\textit{MMF}) allocation can be found with the \textit{water-filling}  algorithm~\cite{Radunovic07}. The  algorithm starts with all path flows from zero. Path flows are then increased equally for all paths, until one or more edges are saturated. We say that such edges are \textit{bottlenecks}, and paths that pass through bottlenecks are also said to be saturated. The path flow of saturated paths is the Max-Min Fair flow allocation of the path. The algorithm is repeated until all paths are saturated, that is until each path passes through at least one bottleneck edge.

We give a visual representation of the water-filling algorithm in Fig.~\ref{fig:Panels}, where a water network connects \textit{taps} (sources) to \textit{buckets} (sinks) with individual paths on  pipelines of unit capacity. Here, the  algorithm proceeds by simultaneously opening the taps at the sources (see Fig.~\ref{fig:Panels}a). In the first iteration of the water-filling  algorithm all paths get an equal flow of $1/3$ and the edge $e_{1,4}$ becomes saturated (see Fig.~\ref{fig:Panels}b). In the second iteration of the  algorithm, the tap at node $1$ cannot be opened any further, as the unit capacity of $e_{1,4}$ has been fully used. Nevertheless, the taps at nodes $2$ and $3$ can still be opened further, and the path flows of paths passing through $e_{2,4}$ and $e_{3,5}$ are increased to $2/3$ (see Fig.~\ref{fig:Panels}c). After the second iteration, the tap at node $2$ cannot be opened further as the edge $e_{4,5}$ is  saturated. On the third iteration, the tap at node $3$ can be opened further, until the path flow on the path passing through $e_{3,5}$ is $1$ (see Fig.~\ref{fig:Panels}d). The  algorithm then stops and the allocation is Max-Min Fair because no path flow can be increased without decreasing another path flow less or equal to it. The analogy with cake-cutting can now be established: at each iteration of the algorithm, the available capacity on the bottleneck edge is shared equally among the paths that cross it and that have not been allocated a share of any edge capacity in previous iterations. In other words, allocating the network capacity can be seen as cutting a sequence of `cakes' (the capacity of bottleneck edges), and the Max-Min Fair allocation establishes the sequence of `cakes' to be shared equally among the paths that cross the bottleneck edges. We are now ready to formalise the iterative MMF  algorithm~\cite{Bertsekas92,Srikant03,Pioro04}. To do this, we first need some mathematical definitions.

\subsection{The Max-Min Fair algorithm explained formally}
\label{subsec:mmf_ algorithm}
 
Let $G=(V,E,c,S,T)$ be a connected and undirected network with node-set
$V$, edge-set $E$, an edge capacity function $c:E\rightarrow
\mathbb{R}_0^{+}$, and a set of source and sink pairs $(s_i,t_i)\in S\times T$ with $i=1,\cdots,M$. Each source and sink pair $(s_i,t_i)$ is connected by a set of $s_i-t_i$ paths $P_i=\{p_{(i,1)},\cdots,p_{(i,k_{i})}\}$, where $P=\cup_{i=1}^{M}P_{i}$ is the set of all source to sink paths on the network. The undirected edge $e_{i,j}\in E$ connects nodes $i$ and $j$, and we will omit the subscript indexes when we are not referring to a specific edge.
All edges $e\in E(p_{(i,k)})$ of a $s_i-t_i$ path $p_{(i,k)}$ 
($1\leq i\leq M$ and $1\leq k\leq  k_{i}$) transport the same \textit{path flow}
$f_{p_{(i,k)}}\in\mathbb{R}^{+}_0$ between $s_i$ and $t_i$. This implies conservation of the flow at the nodes. Different paths can share an edge, even to perform transport in different directions (\eg, during distinct time intervals) \footnote{An alternative would be to consider a directed network, and path flows as a snapshot of network usage in time, but we will not explore this possibility here because undirected networks are traditionally studied before directed ones.}.
The set of paths passing through edge $e\in E$ is  $ P(e)=\left\{ p_{(i,k)}\in  P:e\in E(p_{(i,k)})\right\}$. The flow on edge $e\in E$ is the sum of the path flows on paths passing through the edge, and is given by $F(e)=\sum_{p_{(i,k)}\in P}\delta_{p_{(i,k)}}(e)f_{p_{(i,k)}}$,
where $\delta_{p_{(i,k)}}(e)$ is $1$ if $e\in E(p_{(i,k)})$ and is $0$ otherwise.
A flow is then \textit{feasible} if  $0\leq F(e)\leq c(e)$ for all $e\in E$, where $c(e)$ is the capacity of edge $e$. 
Formally, a set $\{f\}$ of path flows on $ P$ is Max-Min Fair, if it is feasible and if for any other feasible set $\{f^\prime\}$ of path flows on $ P$, there exists a path  $p_{(i,k)}\in P:
f_{p_{(i,k)}}^{\prime}>f_{p_{(i,k)}}$ implies that there exists another path
$p_{(j,l)}\in P:f_{p_{(j,l)}}^\prime< f_{p_{(j,l)}}$ and $f_{p_{(j,l)}}\leq f_{p_{(i,k)}}$.

We define $P^{(i)}$ to be the set of paths on the network at iteration $i$ of the MMF algorithm, and $P^{(i)}(e)$ to be the subset of paths in $P^{(i)}$ that pass through edge $e$.
Before we start the algorithm, we assign $P^{(1)}=P$ and $c^{(1)}(e)=c(e)$ for all $e\in E$, and a path flow $f^{(0)}_{p_{(j,k)}}=0$ to each path $p_{(j,k)}\in P^{(1)}$. We then start the algorithm and initialize the iteration counter $i=1$. 

In the first step of the MMF  algorithm, for each edge $e$ with non-zero capacity that belongs to at least one path, we define the edge capacity divided equally among all paths that cross the edge at iteration $i$ of the MMF  algorithm as
\begin{equation}
\phi^{(i)}(e)=c^{(i)}(e)/\left\vert P^{(i)}(e)\right\vert,
\label{eq:phi}
\end{equation}
for all $c^{(i)}(e)\neq 0$. We then find the minimum of $\phi^{(i)}(e)$, given by
\begin{equation}
\Delta f^{(i)}=\min_{e\in E,c^{(i)}(e)\neq 0}\phi^{(i)}(e).
\label{eq:path_flow_at_step_j}
\end{equation}
In the second step of the MMF  algorithm, we increase all path flows of paths in $P^{(i)}$ by $\Delta f^{(i)}$, such that
\begin{equation}
f_{p_{(j,k)}}^{(i)}=\left\{
\begin{array}
[c]{ll}
f_{p_{(j,k)}}^{(i-1)}+\Delta f^{(i)} & \text{if }p_{(j,k)}\in P^{(i)}\\
f_{p_{(j,k)}}^{(i-1)} & \text{if }p_{(j,k)}\in P\backslash P^{(i)}%
\end{array}
\right.  .
\end{equation}
The effect is to saturate the set of bottleneck edges $E_B^{(i)} =\{e_B \in E:\sum_{p_{(j,k)}\in P^{(i)}}\delta_{p_{(j,k)}}(e_B) \Delta f^{(i)}=c^{(i)}(e_B)\}$, and consequently also to saturate the set of paths that contain at least one bottleneck edge. Next, we create a residual network, by subtracting the capacity used by the path flows,
\begin{equation}
c^{(i+1)}(e)=c^{(i)}(e)-\sum_{p_{(j,k)}\in P^{(i)}}\delta_{p_{(j,k)}}(e)\Delta f^{(i)}.
\label{eq:capacity_update}
\end{equation}
Note that all edges $e_B\in E_B^{(i)}$ will be saturated, that is each will have $c^{(i+1)}(e_B)=0$ after this step. We also say that all paths that contain at least one edge $e_B\in E_B^{(i)}$ are saturated paths, to mean that their path flow will not be increased in subsequent iterations of the MMF algorithm. 
Next, we remove the set of saturated paths from $ P$, that is 
\begin{equation}
P^{(i+1)}   = P^{(i)}\backslash \cup_{e_{B}\in E_{B}^{(i)}}P^{(i)}(e_{B}).
\label{eq:paths_update}
\end{equation}

We say that $P^{(i+1)}$ is the set of augmenting paths because the path flows of paths in $P^{(i+1)}$ can still be increased in subsequent iterations of the algorithm.
If $P^{(i+1)}$ is not empty, we increase the iteration counter, $i\rightarrow i+1$, and we go back to the first step, otherwise we stop and store the value of $i$ as $i^{\ast}$. 
The Max-Min Fair flow on edge $e$ is then the sum of path flows over all paths that cross the edge after the algorithm terminates:
\begin{equation}
F (e)=\sum_{p_{(j,k)}\in P}\delta_{p_{(j,k)}}(e)f_{p_{(j,k)}}^{(i^{\ast})}.
\label{eq:total_flow_final}
\end{equation}
At each iteration, the MMF algorithm increases all path flows equally. Thus, the algorithm finds the bottleneck edges, which determine the maximum path flow possible for path flows that are the smallest. In other words, by finding the minimum of Eq.~(\ref{eq:phi}) at each iteration, we ensure that the smallest path flows are maximized.

\begin{figure}[phtb]
\includegraphics[width=0.48\textwidth]{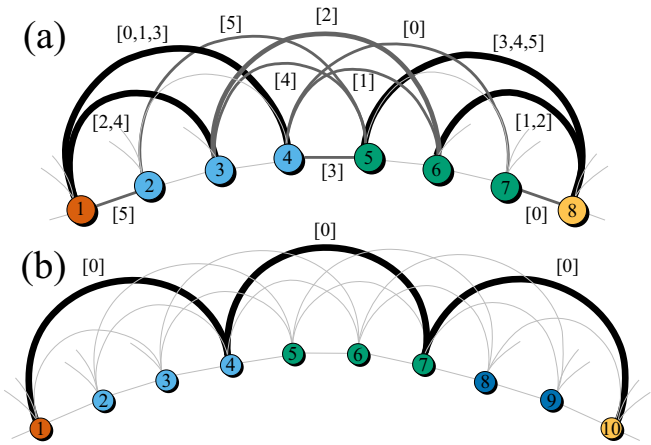}
\caption{\label{fig:visualization_all_panels}(Colour online). Two layouts of $6$-nearest neighbour networks. Nodes are numbered by an index from west to east, and the distance $d$ that separates two nodes is the difference between their indexes. A source $s$ and a sink $t$ pair are placed (a) $d=7$ and (b) $d=9$ nodes apart, respectively. The shortest paths between $s$ and $t$ are identified by numeric labels on the edges. The shortest path length between the two nodes  is $L=3$ in both panels. There are six \textit{s-t} shortest paths in (a), but only one in (b). Edges have unit capacity, edge thickness is proportional to the MMF flow allocation, and saturated edges (or bottlenecks) are drawn in black. Nodes are coloured according to their shortest path length from the source.
}
\end{figure}
\section{Max-Min Fair flows in nearest neighbour networks}
\label{sec:section_III}

\subsection{The network constraints}
\label{subsec:net_constraints}
Recall from Eq.~(\ref{eq:phi}) that the capacity available on the bottleneck edges at each iteration of the MMF algorithm is shared by the paths passing through those edges. Thus, on a network with uniform capacity, the MMF flow allocation depends exclusively on the number of paths passing through each edge. 
To gain a better insight into the MMF allocation of flows, we choose to study $K$-nearest neighbour networks where each edge has constant capacity before the MMF algorithm is applied, that is,
\begin{equation}
c(e)=c^{(1)}(e)=c,\text{ for all }e\in E.
\label{eq:initial_edge_capacity}
\end{equation}
The advantage of studying these regular networks with constant capacity is that we can find simple closed form expressions for counting paths. In these regular graphs of even degree $K$, the nodes are placed on a one dimensional lattice with periodic boundary conditions (the lattice is a ring). Nodes are then connected to their nearest, next-nearest neighbours and so on, up to the constant range $K/2$. $K$-nearest neighbour graphs were recently the object of intense study as prototypes of networks where small amounts of rewiring can generate small-world networks~\cite{Watts98,Newman2010}. To simplify the analysis, we also introduce the constraint that transport only occurs along the set of shortest paths connecting the source and sink nodes.

\begin{figure}[phtb]
\includegraphics[width=0.48\textwidth]{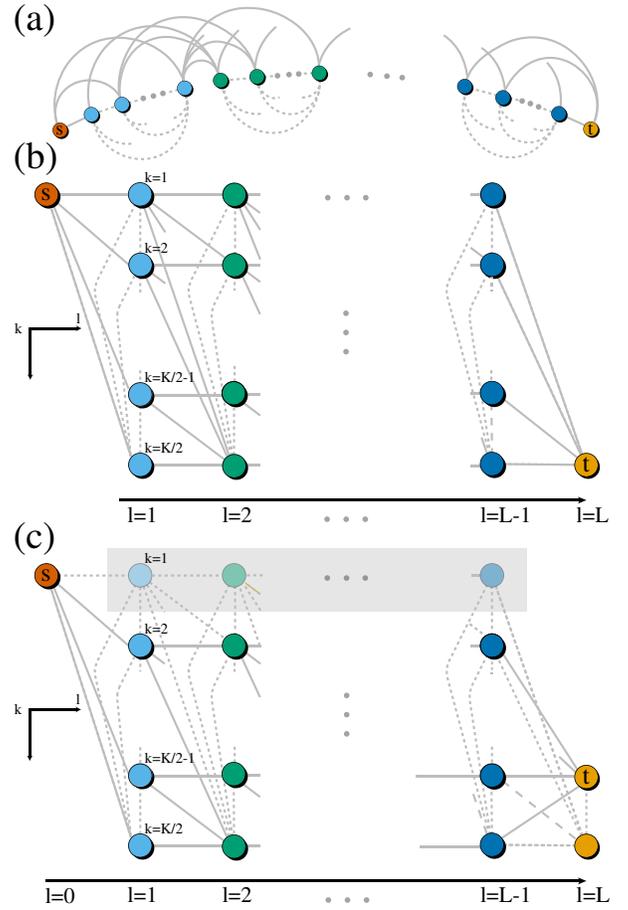}
\caption{\label{fig:scheme}(Colour online). (a) Layout of a $K$-nearest neighbour network where the \textit{s-t} pair is separated by $d^{\min} (K,L)$ nodes on the ring for a given shortest path length $L$. Nodes are coloured according to their shortest path length from the source. Dashed edges connect two nodes at the same shortest path length from the source. (b) Rectangular layout of the network shown in (a), where all nodes except the source and sink are rearranged onto a rectangular lattice $K/2\times (L-1)$ according to their distance from the source. Nodes on this rectangular lattice are identified by their $(k,l)$ coordinates. (c) $K$-nearest neighbour network, where \textit{s-t} are separated by $d = d^{\min} (K,L)+i$ nodes on the ring ($i=1,\dotsc,K/2-1$), has the same structure of shortest paths as a less dense $(K-2i)$-nearest neighbour network, where the two nodes are at the smaller distance $d=d^{\min} (K-2i,L)$. This panel illustrates the case $i=1$, for which the gray area shows that \textit{s-t} shortest paths no longer cross row $k=1$.}
\end{figure}

\subsection{Shortest paths in \texorpdfstring{$K$}{K}-nearest neighbour networks with one (s,t) pair}
\label{sec:shortest_paths_WS}

We place a source and sink pair $(s,t)$ on a $K$-nearest neighbour network with $\vert V\vert $ nodes. Nodes are numbered by an integer index, which indicates their relative position $d\leq  \lfloor \vert V\vert /2\rfloor$ from the source.
Hence, there are two ways of measuring the distance between the source $s$ and the sink $t$. We can either measure the \textit{s-t} distance as the difference $d$ between the indexes of $s$ and $t$ on the ring, or as the shortest path distance $L$ between $s$ and $t$. Since each node has $K$ neighbours, there are $K/2$ nodes at the shortest path distance $L$ from the source, each of them at a distinct distance $d$ from the source. 
Fig.~\ref{fig:visualization_all_panels} is an illustration of the diversity of source to sink (\textit{s-t}) shortest paths when $K=6$ and $L=3$, for $d=7$ and $d=9$. 

\begin{figure}[phtb]
\includegraphics[width=0.48\textwidth]{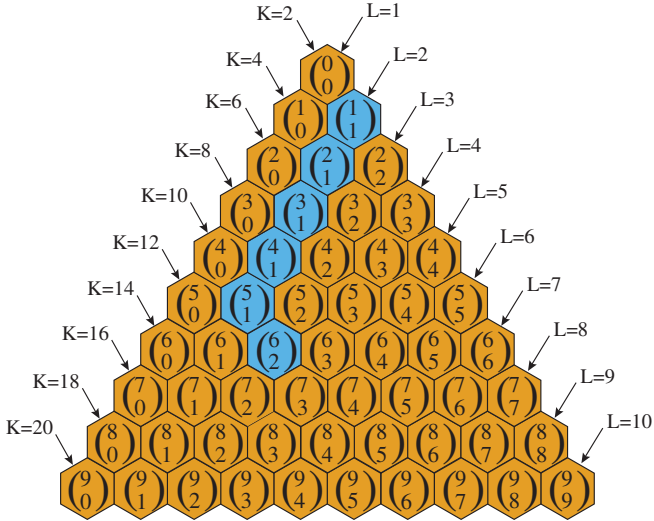}
\caption{\label{fig:pascal}(Colour online). The entries in Pascal's triangle are the number of \textit{s-t} shortest paths between a source and sink nodes at distance $d^{\min} (K,L)$ from each other on a $K$-nearest neighbour network, for a chosen shortest path length $L$ between the two nodes.
Entries of Pascal's triangle coloured in blue illustrate that $\binom{6}{2}$ is the sum of the number of \textit{s-t} shortest paths over the \textit{hockey stick}.}
\end{figure}

We can now analyse the smallest distance $d$ between $s$ and $t$  for a given shortest path length $L$. This distance is given by
\begin{eqnarray}
d^{\min} (K,d)&=&\left[d - \left((d-1) \pmod {K/2} \right)\right]  \label{eq:d_star_K,d} \\
d^{\min} (K,L)&=&(L-1)*K/2+1  \label{eq:d_star},
\end{eqnarray}
as is illustrated in Fig.~\ref{fig:scheme}a. To simplify the counting, we rearrange the network layout as in Fig.~\ref{fig:scheme}b, where each node (except for the source and sink) is identified by its row and column coordinates $(k,l)$, $1\leq  k \leq K/2$ and $1\leq l \leq L-1$. A \textit{s-t} shortest path can be visualised on such \textit{rectangular subgraphs} as a sequence of nodes $s,v_{(k_1,1)},v_{(k_2,2)},\dotsm v_{(k_{L-1},L-1)},t$, where the nodes $v_{(k_i,i)}\in V$ are such that there is an edge between $v_{(k_i,i)}$ and $v_{(k_{i+1},i+1)}$, and $k_1\leq  k_2\leq \dotsb \leq  k_{L-1}$ is a non-decreasing sequence of rows in the rectangular subgraph. In other words, shortest paths contain only  horizontal (east) and diagonal (south east) edges. 
The number of \textit{s-t} shortest paths when the pair is at the distance $d^{\min} (K,L)$ is the number of ways we can distribute $L-1$ identical edges to $K/2$ distinguishable rows. Such allocation can be made in
\begin{equation}
\label{eq:N_beginning_seq}
 N(K/2,L) = \left(\binom{K/2}{L-1}\right) = \binom{K/2+L-2}{L-1}
\end{equation}
different ways, where $\left(\binom{n}{m}\right)=\binom{n+m-1}{m}$ is the number of ways that $m$ indistinguishable balls can be assigned to $n$ distinguishable urns (see the twelvefold way of combinatorics in \cite[pp. 31-38]{RichardStanley00}).
Equation (\ref{eq:N_beginning_seq}) implies that the number of \textit{s-t} shortest paths between a pair of nodes at distance $d^{\min} (K,L)$ is given by an entry in Pascal's triangle as visualised in Fig.~\ref{fig:pascal}.

\begin{figure}[phtb]
\begin{align*}
&\{n(K=2,d)\}_{d=1,\dotsc,\lfloor \vert V\vert /2 \rfloor} = \underbrace{\textcolor{vermilion}{1}}_{L=1},\underbrace{\textcolor{black}{\colorbox{skyblue}{1}}}_{2},\underbrace{\textcolor{vermilion}{1}}_{3}, \underbrace{\textcolor{black}{1}}_{4}, \underbrace{\textcolor{vermilion}{1}}_{5} \dotsc \\
&\{n(K=4,d)\}_{d=1,\dotsc,\lfloor \vert V\vert /2 \rfloor} = \underbrace{\textcolor{vermilion}{1, 1}}_{L=1},\underbrace{\textcolor{black}{\colorbox{skyblue}{2},1}}_{2},\underbrace{\textcolor{vermilion}{3, 1}}_{3}, \underbrace{\textcolor{black}{4, 1}}_{4}, \underbrace{\textcolor{vermilion}{5, 1}}_{5} \dotsc \\
&\{n(K=6,d)\}_{d=1,\dotsc,\lfloor \vert V\vert /2 \rfloor} =  \underbrace{\textcolor{vermilion}{1, 1, 1}}_{L=1},\underbrace{\textcolor{black}{\colorbox{skyblue}{3}, 2,1}}_{2},\underbrace{\textcolor{vermilion}{6, 3, 1}}_{3}, \underbrace{\textcolor{black}{\colorbox{gray}{10}, 4, 1}}_{4},\\ &\underbrace{\textcolor{vermilion}{15, 5, 1}}_{5},\dotsc \\
&\{n(K=8,d)\}_{d=1,\dotsc,\lfloor \vert V\vert /2 \rfloor} =  \underbrace{\textcolor{vermilion}{1, 1, 1, 1}}_{L=1},\underbrace{\textcolor{black}{\colorbox{skyblue}{4}, 3, 2,1}}_{2},\underbrace{\textcolor{vermilion}{10, 6, 3, 1}}_{3},\\ &\underbrace{\textcolor{black}{20, \colorbox{gray}{10}, 4, 1}}_{4},
\underbrace{\textcolor{vermilion}{35, 15, 5, 1}}_{5},\dotsc \\
&\{n(K=10,d)\}_{d=1,\dotsc,\lfloor \vert V\vert /2 \rfloor} =  \underbrace{\textcolor{vermilion}{1, 1, 1, 1, 1}}_{L=1},\underbrace{\textcolor{black}{\colorbox{skyblue}{5}, 4, 3, 2, 1}}_{2},\\ 
&\underbrace{\textcolor{vermilion}{\colorbox{skyblue}{15}, 10, 6, 3, 1}}_{3}, \underbrace{\textcolor{black}{35, 20, 10, 4, 1}}_{4},  \underbrace{\textcolor{vermilion}{70, 35, 15, 5, 1}}_{5},\dotsc 
\end{align*}
\caption{\label{fig:sequences}(Colour online). Batch of sequences denoting the number of shortest paths for a single $(s,t)$ pair positioned on a $K$-nearest neighbour network at distance $d$ from each other. $K$ is the node degree and $L$ is the length of a shortest path between $s$ and $t$. The position of the number of shortest paths for each $K$ value, is the \textit{s-t} distance $d$, \eg~there are $4$ \textit{s-t} shortest paths for $K=8$ at a \textit{s-t} distance $d=5$ and shortest path distance $L=2$ (see node \colorbox{skyblue}{4}). The boxes in blue and gray show that the structure of s-t shortest paths for given $K$ and $d$ can be recovered as a function of less dense K-nearest neighbour networks, where the s-t pair is closer (see also Fig.~\ref{fig:pascal}).}
\end{figure}

We are now able to determine the number of \textit{s-t} shortest paths when the pair is at distance $d^{\min} (K,L)$. Nevertheless, this is not the general case. To illustrate the point, the number of \textit{s-t} shortest paths in nearest neighbour lattices is represented in Fig.~\ref{fig:sequences} as a sequence $\{n(K,d)\}_{d=1,\dotsc,\lfloor \vert V\vert /2 \rfloor}$, which is in itself a batch of monotonically decreasing sequences. Each integer at a position $d$ of the sequence is the number of \textit{s-t} shortest paths when the source and sink nodes are at distance $d$ on the ring. Each of the sub-sequences is characterised by a $(K,L)$ pair, but so far we have only determined the first element of such sub-sequences, which is $N(K/2,d^{\min} )$. These first elements of each sub-sequence follow a \textit{hockey stick} pattern in Pascal's triangle, which is illustrated in Figs.~\ref{fig:pascal} and \ref{fig:sequences} for $K=10$ and $L=3$. Such a pattern tells us that the first element of each subsequence can be found from the sum of previous first elements for a smaller shortest path length $L-1$, that is $N(K/2,L)=\sum_{i=1}^{K/2}N(i,L-1)$.

To understand the structure of the sub-sequences, observe that the set of shortest paths on a $K$-nearest neighbour network when $s$ and $t$ are placed $d=d^{\min}(K,L)+i$ nodes apart is equivalent to the set of $s-t$ shortest paths on a less dense $(K-2i)$-nearest neighbour network where $s$ and $t$ are placed closer to each other, at the distance $d^{\min}(K-2i,L)$. 
For example, if $K=8$ and $L=4$, and the \textit{s-t} nodes are at distance $d^{\min}(8,4)=13$, increasing the distance to $d=14$ does not change the shortest path length $L$, but the set of shortest paths between the two nodes is  the same as when the $\textit{s-t}$ pair are placed on a $6$-nearest neighbour network at distance $d^{\min}(6,4)=10$.
This effect is illustrated in Fig. ~\ref{fig:sequences} (see the nodes in the gray boxes) and in Figs.~\ref{fig:scheme}b and c for $i=1$.
As can be visualised from Fig.~\ref{fig:scheme}c, increasing the lattice position of the sink by one has the consequence that the connections between the new sink node and nodes $(k=1,l)$ for $1 \leq l \leq L-1$ (the first row in the rectangular subgraph of Fig.~\ref{fig:scheme}c) no longer belong to any \textit{s-t} shortest paths. The argument can be easily generalized to any $i$. 

\begin{figure}[phtb]
\includegraphics[width=0.48\textwidth]{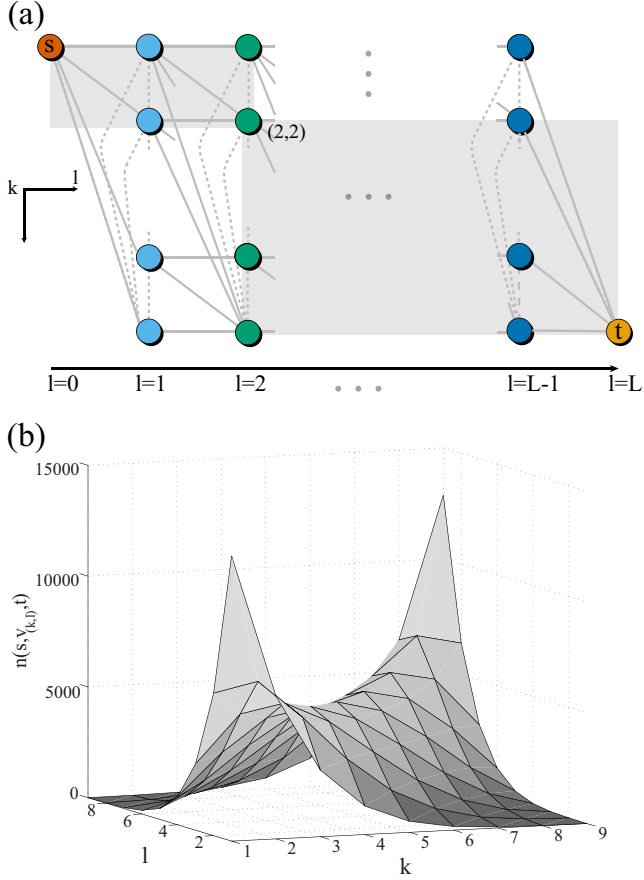}
\caption{\label{fig:n_paths_through_a_point}(Colour online). (a) The number $n(s,v_{(2,2)},t)$ of \textit{s-t} shortest paths that cross node $v_{(2,2)}$ is determined from the product rule as the product of the number of \textit{s-t} shortest paths in the two rectangular subgraphs highlighted in gray. (b) The number $n(s,v_{(k,l)},t)$ of \textit{s-t} shortest paths  between the source $s$ and the sink $t$ through the node $(k,l)$ (where $L=10$ and $K=12$) has a saddle shape, with maxima at $(k=1,l=1)$ and $(k=K/2,l=L-1)$.}
\end{figure}

The number of \textit{s-t} shortest paths passing through a set of given nodes is given by the product rule.
The number of shortest paths connecting nodes $s$ and $t$ and passing through the node $v_{(k,l)}$ in the rectangular subgraph is given by:

\begin{align}
\label{eq:N_K_l_m_l}
n(s,v_{(k,l)},t) &= N(k,l)N(K/2-k+1,L-l) \nonumber \\  
&= \left(\binom{k}{l-1}\right)\left(\binom{K/2-k+1}{L-l-1}\right), 
\end{align}
where  $1 \leq  k \leq K/2$ and $1 \leq l \leq L-1$. Figure~\ref{fig:n_paths_through_a_point}a shows schematically how Eq.~(\ref{eq:N_K_l_m_l}) is derived, and  Fig.~\ref{fig:n_paths_through_a_point}b is a plot of $n(s,v_{(k,l)},t)$ that illustrates that the concentration of shortest paths is higher along the diagonal of the rectangular layout. The number of shortest paths connecting nodes $s = (k_0,l_0) = (1,0)$ and $t = (k_{n+1},l_{n+1}) = (K/2, L)$ and passing through nodes $v_{(k_1,l_1)},\dotsc,v_{(k_n,l_n)}$ is given by:

\begin{align}
\label{eq:N_K_l_m1_l1_mn_ln} 
&n(s,v_{(k_1,l_1)},\dotsc,v_{(k_n,l_n)},t)   \nonumber \\ 
&= \prod\limits_{i=0}^{n} N(k_{i+1}-k_{i}+1,l_{i+1}-l_{i})  \nonumber \\ 
&= \prod\limits_{i=0}^{n}\left(\binom{k_{i+1}-k_{i}+1}{l_{i+1}-l_{i}-1}\right),
\end{align}
where $1 \leq  k_1 \leq \dotsb \leq  k_{n} \leq K/2$ and $1 
\leq l_1 < l_2 < \dotsb < l_n \leq L-1$. From Eq. (\ref{eq:N_K_l_m_l}), the number of shortest paths containing edge $e_{s,(k,1)}$ is defined by $n(e_{s,v_{(k,1)}}) = n(s,v_{(k,1)},t)$ and is given by
\begin{align}
\label{eq:N_K_l_m1_1}
 n(e_{s,v_{(k,1)}}) &= N(K/2-k+1,L-1) \nonumber \\
&= \left(\binom{K/2-k+1}{L-2}\right).
\end{align}
Similarly, from Eq.~(\ref{eq:N_K_l_m1_l1_mn_ln}) the number of \text{s-t} shortest paths passing through a set of nodes, we derive the 
number of \text{s-t} shortest paths containing edges $e_{s,(k_1,1)}$ and $e_{(k_2,L-1),t}$ that is defined as $n(e_{s,v_{(k_1,1)}},e_{v_{(k_2,L-1)},t}) = n(s,v_{(k,1)},v_{(k_2,L-1)},t)$, to be 
\begin{align}
\label{eq:N_K_l_m1_1_m2_(lst-1)}
n(e_{s,v_{(k_1,1)}},e_{v_{(k_2,L-1)},t}) &=  N(k_2-k_1+1, L-2) \nonumber \\ 
&=  \left(\binom{k_2-k_1+1}{L-3}\right).
\end{align}
Equations (\ref{eq:N_K_l_m_l}) to (\ref{eq:N_K_l_m1_1_m2_(lst-1)}) show that the number of \textit{s-t} shortest paths passing through a node or a set of nodes can be written as a product of binomials. Each of these binomials is the number of \textit{s-t} shortest paths when the \textit{s-t} pair is located at the distance $d^{\min} $ on networks with specific $K$ and $L$ values and is given by Eq.~(\ref{eq:N_beginning_seq}).

To summarise, the number of \textit{s-t} shortest paths depends on $d$ and is given by a batch of sequences (see Fig.~\ref{fig:sequences}), each having $K/2$ elements. If the \textit{s-t} pair is at the smallest distance $d^{\min} (K,L)$ given by Eq.~(\ref{eq:d_star}) for a given shortest path length $L$, then the number of \textit{s-t} shortest paths is maximal for that $L$. Increasing the distance between the \textit{s-t} pair to $d=d^{\min} (K,L)+i$ (for $1\leq i\leq K/2-1$) creates the same structure of \textit{s-t} shortest paths as placing the \textit{s-t} pair on a $(K-2i)$-nearest neighbour network at the distance $d^{\min} (K-2i,L)$, which is the smallest distance at which the two nodes can be located on a $(K-2i)$-nearest neighbour network so that the shortest path length between them is $L$. This implies that the problem of determining sink inflow on a $K$-nearest neighbour network, as the distance between the \textit{s-t} pair is varied, is reduced to the problem of calculating sink inflow at specific distances $d^{\min} (K,L)$ given by Eq.~(\ref{eq:d_star}) as $K$ and $L$ are varied. Hence, we concentrate on determining the sink inflow rigorously for \textit{s-t} pairs placed at distances $\{d^{\min}(K,L)\}\vert  _{L=1,\dotsc,\infty}$ on $K$-nearest neighbour networks. From now on we will refer to a \textit{s-t} pair at the shortest path length $L$ to mean that the \textit{s-t} pair is at the distance $d=d^{\min}(K,L)$, where $d^{\min}(K,L)$ is related to $L$ in Eq.~(\ref{eq:d_star}).
As we will see in Sec. \ref{sec:paths_counting}, the calculation of the sink inflow and of the path flows requires the number of \textit{s-t} shortest paths in various situations, as given by Eqs.~(\ref{eq:N_K_l_m_l}) to (\ref{eq:N_K_l_m1_1_m2_(lst-1)}).

\subsection{Exact calculation of flows with the path counting methods in 
\texorpdfstring{$K$}{K}-nearest neighbour networks with one (s,t) pair}
\label{sec:paths_counting}

\begin{figure}[phtb]
\includegraphics[width=0.48\textwidth]{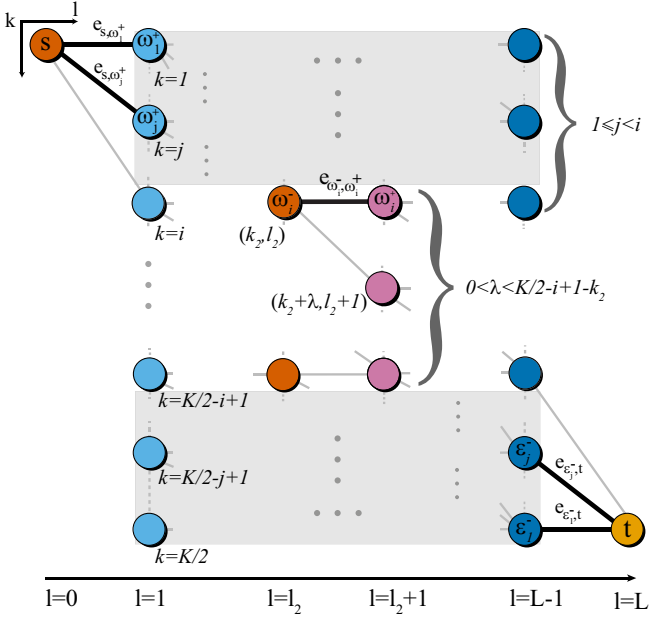}
\caption{\label{fig:simplifying_scheme}(Colour online). Rectangular layout of the subgraph between the source ($s$) and the sink ($t$) nodes at iteration $i$ of the MMF algorithm. The bottlenecks found at iterations $1\leq j<i$ are edges of $s$ or $t$ and are represented as thick black edges. The pair of bottlenecks found at iteration $i$ of the MMF algorithm, are identified by the notation $e_{\omega _i^-,\omega _i^+}$ (the western bottleneck) and $e_{\varepsilon _i^-,\varepsilon _i^+}$ (the eastern bottleneck, not represented). When the bottlenecks are edges of the source or sink, $\omega _i^-=s$ and $\varepsilon _i^+=t$, respectively.}
\end{figure}

Our goal is now to get analytical insights into the MMF allocation on $K$-nearest neighbour networks with one \textit{s-t} pair. In particular, we wish to determine all path flows as a function of the distance $L$ between the source and sink nodes. The knowledge of all path flows also allows us to determine the incoming flow at the sink node as a function of $L$. The sink inflow can be determined as the sum of the path flows over the $K/2$ edges of the sink that contain shortest paths,
\begin{equation}
\label{eq:sink_inflow}
F(K,L)=\sum_{i=1}^{K/2}\sum_{p(j,k)\in P}\delta_{p(j,k)}(e_{v_i,t})f_{p(j,k)}^{(i^{\ast})},
\end{equation}
where $v_{i}=(K/2-i+1,L-1)$, $i=1,\dotsc,K/2$ are the $K/2$ neighbours of the sink that belong to \textit{s-t} shortest paths.

Such a problem becomes more interesting as we realise that the contribution to the sink inflow from each iteration of the MMF algorithm can be found by path counting methods alone. Indeed, the ratio of capacity to the number of paths for an edge on Eq.~(\ref{eq:phi}), is a function of both the edge capacity on the residual network and the number of unsaturated paths passing through the edge. In turn, the residual edge capacity depends exclusively on the number of paths that pass through the edge and are saturated at the previous iterations of the MMF algorithm. This implies that the calculation of sink inflow can be done rigorously from path counting alone, once we have found the pattern of the location of bottleneck edges at each iteration of the MMF algorithm. 

We start by assuming that all bottlenecks are found on edges of the source and sink nodes (see Fig. \ref{fig:simplifying_scheme}). When $K/2$ is odd, this means that the row $\lceil K/4 \rceil$ in the rectangular layout contains two bottleneck edges instead of one, and we observe a total of $2\lceil K/4 \rceil$ bottlenecks. The network is symmetric (\ie, the source and sink can be interchanged) and thus bottlenecks are positioned symmetrically in pairs. We introduce the notation $e_{\omega _i^-,\omega _i^+}$ and $e_{\varepsilon _i^-,\varepsilon _i^+}$, to refer to a symmetric pair of bottleneck edges found at iteration $i$. In general, we will refer to a bottleneck closer to the source $e_{\omega _i^-,\omega _i^+}$ as a \textit{western} bottleneck and a bottleneck closer to the sink $e_{\varepsilon _i^-,\varepsilon _i^+}$ as an \textit{eastern} bottleneck. Thus, at the $i$-th iteration of the MMF  algorithm, the western bottleneck  $e_{s,\omega _i^+}$ connects the source and node 
\begin{equation}
\omega _i^+=v_{(i,1)},
\label{eq:b_m}
\end{equation}
and a symmetric eastern bottleneck $e_{{\varepsilon _i^-},t}$ will be located between the sink and node 
\begin{equation}
{\varepsilon _i^-}=v_{(K/2-i+1,L-1)}.
\label{eq:b_m_prime}
\end{equation}

Since the bottleneck edges found at iteration $i$ are located on $e_{s,\omega _i^+}$ and the symmetric edge $e_{{\varepsilon _i^-},t}$, the path flow increment (\ref{eq:path_flow_at_step_j}) at iteration $i$ of the MMF algorithm is given by 
\begin{align}
\Delta f^{(i)} & =\min_{e\in E}\phi^{(i)}(e)=\phi^{(i)}(e_{s,\omega _i^+})\nonumber \\
& =c^{(i)}(e_{s,\omega _i^+})/\left\vert P^{(i)}(e_{s,\omega _i^+})\right\vert,
\label{eq:Delta_f}
\end{align}
for all $c^{(i)}(e_{s,\omega _i^+})\neq 0$.
While the denominator of (\ref{eq:Delta_f}) depends only on the number of unsaturated \textit{s-t} shortest paths passing through $e_{s,\omega _i^+}$ at iteration $i$, the residual capacity $c^{(i)}(e_{s,\omega _i^+})$ on the numerator is found recursively as a function of the path flow increment added to unsaturated paths at iterations $j<i$ of the MMF algorithm. This recursive calculation depends on the number $\left\vert P^{(j)}(e_{s,\omega _i^+})\right\vert$ of \textit{s-t} shortest paths passing through edge $e_{s,\omega _i^+}$ that have not been saturated until iteration  $j<i$. The path flow of each of these paths is increased by $\Delta f^{(j)}$ at iteration $j$, and the capacity of all edges that these paths cross is reduced by $\Delta f^{(j)}$.

Paths passing through $e_{s,\omega _i^+}$  that have been saturated in the previous iteration $j<i$ of the MMF  algorithm, have the property that each cross  $e_{s,\omega _i^+}$ as well as one of the bottleneck edges $e_{{\varepsilon _q^-},t}$ found at a previous iteration $q=1,\dotsc,j-1$. 
For each $q=1,\dotsc,j-1$, there are $n(e_{s,\omega _i^+},e_{{\varepsilon _q^-},t})$ such paths given by Eq.~(\ref{eq:N_K_l_m1_1_m2_(lst-1)}). Hence, $\left\vert P^{(j)}(e_{s,\omega _i^+})\right\vert$, which is the number of unsaturated paths at iteration $j$ crossing $e_{s,\omega _i^+}$, can be determined  from Eqs.~(\ref{eq:N_K_l_m1_1}) and (\ref{eq:N_K_l_m1_1_m2_(lst-1)}) as
\begin{align}
\label{eq:eta}
\left\vert P^{(j)}(e_{s,\omega _i^+})\right\vert & = n(e_{s,\omega _i^+})-\sum_{q=1}^{j-1}n(e_{s,\omega _i^+},e_{{\varepsilon _q^-},t}) \nonumber \\
& = n(e_{s,\omega_{(i+j-1)}^{+}})\nonumber \\
& = N(K/2-(i+j)+2,L-1),
\end{align}
where $1\leq i+j-1\leq K/2-1$, and from the paths counting formulas ~(\ref{eq:N_beginning_seq}), (\ref{eq:N_K_l_m1_1}) and (\ref{eq:N_K_l_m1_1_m2_(lst-1)}), and the location (\ref{eq:b_m}) and (\ref{eq:b_m_prime}) of the western and eastern bottlenecks, 
\begin{equation}
\label{eq:N_bi_plus}
n(e_{s,\omega _i^+})=N(K/2-i+1,L-1)
\end{equation}
and 
\begin{equation}
\label{eq:N_bi_bq}
n(e_{s,\omega _i^+},e_{{\varepsilon _q^-},t})=N(q-i+1,L-2).
\end{equation}
To understand Eq.~($\ref{eq:eta}$), observe that the \textit{s-t} shortest paths crossing both $e_{s,\omega _i^+}$ and edges in row $K/2-j+1$ of the rectangular subgraph (see Fig. \ref{fig:scheme}) also cross bottleneck $e_{{\varepsilon _j^-},t}$, and these paths are saturated at iteration $j<i$ of the MMF algorithm. By symmetry, the \textit{s-t} shortest paths crossing $e_{{\varepsilon _i^-},t}$ and edges in row $j$ also cross the bottleneck $e_{s,\omega _j^+}$, \ie~they are also saturated at iteration $j$. The effect of this pattern of saturated \textit{s-t} shortest paths is that, for each iteration $j<i$, the MMF algorithm saturates two bottleneck edges, $e_{s,\omega _j^+}$ and $e_{{\varepsilon _j^-},t}$, as well as all the paths that cross  rows $j$ and $K/2-j+1$ of the rectangular subgraph.
Thus, the effect of $j$ iterations of the MMF algorithm is to remove $2j$ rows ($j$ rows from the top and $j$ rows from the bottom) of the 
original $K/2$-by-$L$ rectangular subgraph, creating an equivalent $(K/2-2j)$-by-$L$ rectangular subgraph. Note that the equivalent rectangular subgraph is only useful for path counting purposes, since the capacity of its edges will have been reduced after $j$ steps of the MMF algorithm.

Each of the $\left\vert P^{(j)}(e_{s,\omega _i^+})\right\vert$ unsaturated paths has its path flow incremented by $\Delta f^{(j)}$ at iteration $j<i$,  according to Eq.~(\ref{eq:Delta_f}). So, each path in $P^{(j)}(e_{s,\omega _i^+})$ reduces the available capacity $c^{(j)}(e_{s,\omega _i^+})$  by $\Delta f^{(j)}$ according to Eq. (\ref{eq:capacity_update}). Since all edges have initial capacity $c$ (see Eq. ~(\ref{eq:initial_edge_capacity})), the path flow increment at iteration $i$ of the MMF algorithm, given by Eq.~(\ref{eq:Delta_f}), can be written as a function of Eq.~(\ref{eq:eta}):
\begin{align}
\label{eq:delta_paths}
\Delta f^{(i)} & = \frac{c-\sum_{q=0}^{i-1}\left\vert P^{(q)}(e_{s,\omega _i^+})\right\vert\Delta f^{(q)}}{\left\vert P^{(i)}(e_{s,\omega _i^+})\right\vert} \nonumber \\
& = \frac{c-\sum_{q=0}^{i-1} n(e_{s,\omega_{(i+q-1)}^+}) \Delta f^{(q)}}{n(e_{s,\omega _{(2i-1)}^+})} \nonumber \\
& = \frac{c-\sum_{q=0}^{i-1} N(K/2-(i+q)+2,L-1) \Delta f^{(q)}}{N(K/2-2(i-1),L-1)},
\end{align}
where $\Delta f^{(0)}=0$. Equation~(\ref{eq:delta_paths}) says that the increment to path flows $\Delta f^{(i)}$ at iteration $i$ on a $K/2$-by-$L$ rectangular subgraph can be written as a recursive function of the number of \textit{s-t} shortest paths on a sequence of nearest-neighbour induced subgraphs with smaller node degree than the original rectangular subgraph.

\begin{figure}[phtb]
\includegraphics[width=0.48\textwidth]{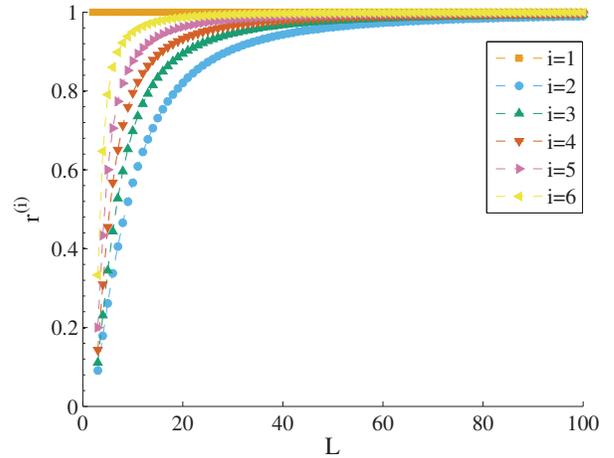}
\caption{\label{fig:r}(Colour online). Plot of the ratio $r^{(i)}$ between the path flow increment at iteration $i$ and the path flow of paths that are saturated at iteration $i$, as a function of $L$ for $K=24$ as the iterations $i$ increase. The path flow is dominated by the path flow increment at the last iteration when $r\simeq1$.}
\end{figure}

To gain insight into the Max-Min Fair path flows $f^{(i^\ast)}$ when all bottlenecks are edges of $s$ or $t$, we plot in Fig.~\ref{fig:r} the ratio
\begin{equation}
\label{eq:r}
r^{(i)}=\frac{\Delta f^{(i)}}{f^{(i^\ast)}}=\frac{\Delta f^{(i)}}{\sum_{j=1}^{i}\Delta f^{(j)}},
\end{equation}
for all paths that are saturated at iteration $i$. These paths that are saturated at iteration $i$ are all the paths that pass through a bottleneck found at iteration $i$, or equivalently, all paths in $P^{(i)}(e_B)$. Thus, the ratio $r^{(i)}$ represents the relative contribution of the path flow increment at iteration $i$ to the path flow assigned to paths that are saturated at iteration $i$.
The observed $r^{(i)}\rightarrow 1$ tells us that the path flows allocated by the MMF algorithm are dominated by the flow increment $\Delta f^{(i)}$ at the last iteration. A histogram of the path flows is thus well approximated by a plot of the number~$\eta^{(i)}=N(K/2-2(i-1),L)-N(K/2-2i,L)$ of paths  saturated at iteration $i$ versus $\Delta f^{(i)}\propto1/N(K/2-2(i-1),L-1)$, and for $L$ large $\Delta f^{(i)}\sim 1/\eta^{(i)}$. In other words, when the bottlenecks are edges of $s$ or $t$ and $L$ is large, the histogram of path flows is a discrete set of 
points on a line with slope $-1$.

Combining the number of \textit{s-t} shortest paths and the path flow increment according to Eqs.~(\ref{eq:N_beginning_seq}) and  (\ref{eq:delta_paths}), the MMF flow at the sink is then 
\begin{equation}
\label{eq:sink_inflow2}
F (K,L)=\sum_{j=1}^{\lceil K/4 \rceil}N(K/2-2(j-1),L)\Delta f^{(j)}
\end{equation}
Combining the number of \textit{s-t} shortest paths Eqs.~(\ref{eq:N_beginning_seq}) and the number (\ref{eq:N_bi_plus}) of paths passing through the western bottleneck, Eq. (\ref{eq:sink_inflow2}) can be simplified for large $L$ to
\begin{equation}
\label{eq:sink_inflow_limit_binomials}
F(K,L)\sim \sum_{j=1}^{\lceil K/4\rceil }\frac{\binom{K/2-2j+L} {L-1}}{\binom{K/2-2j+L-1}{L-2}}c. 
\end{equation}
Using the asymptotic expansion of the binomial coefficient, Eq.~(\ref{eq:sink_inflow_limit_binomials}) can be simplified further to 
\begin{equation}
\label{eq:sink_inflow_expansion}
F(K,L)\sim c\sum_{j=1}^{\lceil K/4\rceil }\frac{(L-1)^{K/2-2j+1}}{(L-2)^{^{K/2-2j+1}}},
\end{equation}
and thus 
\begin{equation}
\label{eq:sink_inflow_limit}
\lim_{L\rightarrow\infty}F(K,L)=\lceil K/4\rceil c.
\end{equation}
This asymptotic behaviour can be easily observed in Table \ref{tab:sink_inflow} for small $K$. 

\begin{table}[phtb]
\caption{\label{tab:sink_inflow}Sink inflow $F(K,L)$ as a function of \textit{s-t} shortest path length $L$ for node degree $4\leq K\leq 14$, when all bottlenecks are edges of $s$ or $t$. The asymptotic behaviour of the sink inflow is $\lim_{L\rightarrow\infty}F (K,L)=\lceil K/4 \rceil c$.}
\begin{ruledtabular}
\begin{tabular}{ll}
K & F(K,L)\\
$4$ & $\frac{Lc}{L-1}$ \\
$6$ & $2c+ \frac{2}{\left(L-1\right) L}c$ \\
$8$ & $2c+ \frac{L+4}{L^2-1}c$ \\
$10$ & $3c+ \frac{2 \left(L^2+11 L-2\right)}{L^4+2 L^3-L^2-2 L}c$ \\

$12$ & $3c+ \frac{L^3+9 L^2+51 L+34}{L^4+5 L^3+5 L^2-5 L-6}c$ \\
$14$ & $4c+ \frac{2 \left(L^4+18 L^3+189 L^2+88 L-12\right) }{L^6+9 L^5+25L^4+15 L^3-26 L^2-24 L}c$ 
\end{tabular}
\end{ruledtabular}
\end{table}

So far, we have made the assumption that, at each iteration of the MMF algorithm, the algorithm saturates one edge of the source and another of the sink, up to a total of $2\lceil K/4 \rceil$ bottleneck edges. This assumption allows us to compute exactly the path flow increments (see Eq.~(\ref{eq:delta_paths})) and the sink inflow (see Eq.~(\ref{eq:sink_inflow2})).
Knowing the pattern of the location of bottleneck edges gives us a clear advantage over solving the MMF allocation with the generic implementation of the algorithm described in Sec.~\ref{subsec:mmf_ algorithm}. Indeed, the generic implementation of the MMF algorithm has to keep track of all paths and path flows at each iteration. This process consumes increasing resources as $K$ and $L$ increase, owing to the number of  \textit{s-t} shortest paths growing polynomially or faster with either $K$ or $L$. Thus, knowledge of the location of the bottleneck edges is particularly important for some regions of the parameter space, where the number of \textit{s-t} shortest paths is large, and it is impractical to implement the MMF algorithm as described in Sec.~\ref{subsec:mmf_ algorithm}. 

Since our assumption has been that the bottlenecks are edges of $s$ or $t$, it is then natural to investigate the regions of the parameter space $(K,L)$ where this assumption holds. To answer this question for each $(K,L)$ pair, we compute the smallest iteration of the MMF algorithm such that the bottleneck edges are not edges of $s$ or $t$. Since the bottleneck edges found at iteration $i$ are located on
rows $i$ and $K/2-i+1$, to find these bottlenecks we need to search for the minimum of Eq.~(\ref{eq:phi}) over all edges that link to a node in row $i$. This search can be greatly simplified when all $2(i-1)$ bottlenecks found up to iteration $i-1$ are edges of $s$ or $t$. 
Let us assume that the western bottleneck $e_{\omega _i^-,\omega _i^+}$, which is not an edge of $s$ or $t$, connects the two nodes
\begin{equation}
\left\{\begin{aligned}
\omega _i^- & =  v_{(k_2,l_2)}\\
\omega _i^+ & =  v_{(k_2+\lambda,l_2+1)},
\end{aligned}
\right.
\label{eq:inner_bottlenecks}
\end{equation}
where $i\leq  k_2\leq \lceil K/4 \rceil$, $0\leq \lambda\leq K/2-i+1-k_2$ is an integer, and the eastern bottleneck is defined in a similar way.
When $\lambda=0$, we say that the bottleneck is \textit{horizontal} because it links two nodes that share the same row and are placed on adjacent columns on the rectangular subgraph in Fig.~\ref{fig:scheme}. When $\lambda\neq 0$, we say that the bottleneck is \textit{diagonal}, because it connects two nodes that belong to adjacent columns, but different rows of the rectangular subgraph (see Fig.~\ref{fig:simplifying_scheme}). 
The search for bottleneck edges is simplified because we only need to search over horizontal edges. The proof consists in showing that  the numerator (denominator) of Eq.~(\ref{eq:phi}) has a minimum (maximum) for horizontal edges, i.e. when $\lambda=0$ and thus the search can be restricted to horizontal edges. 

We start by observing that the numerator of Eq.~(\ref{eq:phi}) is the edge capacity on the residual network at iteration $i$. To show that this edge capacity is the smallest when $e_{\omega_i^-,\omega_i^+}$ is a horizontal edge, we demonstrate that it decreases at iteration $i$ by the largest value when $e_{\omega_i^-,\omega_i^+}$ is horizontal, according to Eq.~(\ref{eq:capacity_update}). Such decrease is given, at each iteration $j<i$ of the MMF algorithm, by multiplying the path flow increment Eq. (\ref{eq:path_flow_at_step_j}) by the number of paths that pass through $e_{\omega_i^-,\omega_i^+}$ and are saturated at that iteration $j$. The path flow increment is constant for all paths at each iteration. So we just need to show that, for all iterations $j<i$, the number of \textit{s-t} shortest paths saturated at each iteration $j$, passing through $e_{\omega_i^-,\omega_i^+}$ is larger when $e_{\omega_i^-,\omega_i^+}$ is a horizontal, than when it is a diagonal edge. Recall that here we are assuming that all bottlenecks found until iteration $i-1$ are edges of $s$ or $t$.

Let us start with a $K/2$-by-$(L-1)$ network. At iteration $j<i$, we have an equivalent $(K/2-2(j-1)))$-by-$(L-1)$ network. The paths that are saturated at iteration $j$ and cross $e_{\omega _i^-,\omega_i^+}$ also cross \textit{i)} the j-th western bottleneck $e_{s,v_{(j,1)}}$ and one of the edges $e_{v_{(K/2-k+1,L-1)},t}$ for $j\leq  k <i$; \textit{ii)} the j-th eastern bottleneck $e_{v_{(K/2-j+1,L-1)},t}$ and one of the edges $e_{s,v_{(k,1)}}$ for $j<  k <i$. Therefore, the total number of paths saturated at iteration $j$ crossing $e_{\omega _i^-,\omega _i^+}$ is

\begin{align}
n_{sat}^{(j)}(e_{\omega _i^-,\omega _i^+})& =  n( e_{s,v_{(j,1)}},e_{\omega _i^-,\omega _i^+},e_{v_{(K/2-j+1,L-1)},t}) \nonumber\\
& + \sum_{k=j+1}^{k_2}n( e_{s,v_{(k,1)}},e_{\omega _i^-,\omega _i^+},e_{v_{(K/2-j+1,L-1)},t}) \nonumber\\
& + \sum_{k=k_2 + \lambda}^{K/2-j}n( e_{s,v_{(j,1)}},e_{\omega _i^-,\omega _i^+},e_{v_{(k,L-1)},t}). 
\label{eq:n_paths_through_inner_bottleneck}
\end{align}
The symmetry of paths from the source to the sink and vice-versa, together with Pascal's rule allow us to simplify Eq.~(\ref{eq:n_paths_through_inner_bottleneck}) to
\begin{align}
&n_{sat}^{(j)}(e_{\omega _i^-,\omega _i^+})) =  \binom{-j+\frac{K}{2}+L-l_2-\lambda -k_2-2}{L-l_2-3} \nonumber\\
& \left[\binom{-j-l_2+k_2-2}{l_2-2} + \binom{-j-l_2+k_2-2}{l_2-1}    \right] \nonumber\\
& + \binom{-j-l_2+k_2-2}{l_2-2} \nonumber\\
& \left[ \binom{-j+\frac{K}{2}+L-l_2-\lambda-k_2-2}{L-l_2-2} - (L-l_2-3)\right].
\label{eq:simplified_n_paths_through_inner_bottleneck}
\end{align}
Observe that $K/2-j-k_2+2\ge 0$, and thus the total number of paths saturated at iteration $j<i$ that cross $e_{\omega _i^-,\omega _i^+}$, which is given by
Eq.~(\ref{eq:simplified_n_paths_through_inner_bottleneck}), is maximum for $\lambda=0$, that is when $e_{\omega _i^-,\omega _i^+}$ is a horizontal edge. This implies that the numerator of Eq.~(\ref{eq:phi}) has a minimum when the edge is horizontal. Next, we show that the denominator of Eq.~(\ref{eq:phi}) has a maximum for horizontal edges. To do this, we need to demonstrate that the number of unsaturated paths crossing $e_{\omega _i^-,\omega _i^+}$ at iteration $j < i$ is larger for horizontal edges than for diagonal ones. Note that all paths passing through the $\left (K/2-2(j-1)\right )$-by-$(L-1)$ residual network are unsaturated at iteration $j$, so the number of unsaturated \textit{s-t} shortest paths that cross $e_{\omega _i^-,\omega _i^+}$ at iteration $j$ is found from Eqs.~(\ref{eq:N_K_l_m1_l1_mn_ln}) and (\ref{eq:inner_bottlenecks}) as
\begin{align}
\label{eq:n_paths_through_4_nodes}
& n_{unsat}^{(j)}(e_{\omega _i^-,\omega _i^+}) = \nonumber\\ & \left ( \binom{k_2-j+1}{l_2-1}\right ) \left (\binom{K/2-j-k_2-\lambda+2}{L-l_2-2}\right ).
\end{align}

Equation~(\ref{eq:n_paths_through_4_nodes}) yields the maximum value for $\lambda=0$, \ie~for horizontal edges. Thus, the minimum of Eq.~(\ref{eq:phi}) is on a horizontal edge. This means that when the first $2(i-1)$ bottlenecks are edges of the source or sink, the $2i$-th bottleneck is either again an edge of $s$ or $t$, or it is a horizontal bottleneck.

Taken together, these results yield a simple procedure to locate the bottlenecks edges. Starting from the first row $i=1$ on the rectangular subgraph, we search for the pair of bottleneck edges, i.e. the minima of Eq.~(\ref{eq:phi}) over the set of horizontal edges on that row. When $i=1$, the bottlenecks are edges of $s$ or $t$. Nevertheless, this is not necessarily the case when $i>1$. In other words, there may exist an iteration $i$ such that the bottlenecks previously found are edges of $s$ or $t$, but the new pair of bottlenecks is not.
We then know the exact location of the $2i$ bottleneck edges found up to iteration $i$. When $i$ is the last iteration of the MMF algorithm, we can derive the Max-Min Fair flows analytically from the path counting methods developed above, otherwise we still have an approximation of the MMF flow that is exact up to iteration $i$ of the algorithm. 
Applying a similar reasoning to the one presented for the first inner bottleneck, it is possible to show analytically that all other bottlenecks, except the bottlenecks of the source (or sink), are horizontal edges as well when \textit{i)} each row $k_j < k_i$ has at least one bottleneck edge found in a previous iteration; \textit{ii)} the last bottleneck to be found at each row $k_j < k_i$ is an edge of the source (or the sink). The first condition guarantees that, up to iteration $i$,  all rows up to row $k_i$ have at least one bottleneck. The second condition guarantees that all \textit{s-t} shortest paths crossing any edge in all rows $k_j < k_i$ have been saturated, and thus the only unsaturated paths are in the $(K/2-2(i-1))$-by-$(L - 1)$ rectangular subgraph. Since these two conditions simplify the path counting,  we are able to compute the flow distribution efficiently by a semi-analytical method. This method relies on an analytical analysis to restrict the search to horizontal links, combined with a numerical search over the same horizontal links. 

\section{Results}
\label{sec:results}

\begin{figure}[phtb]
\includegraphics[width=0.48\textwidth]{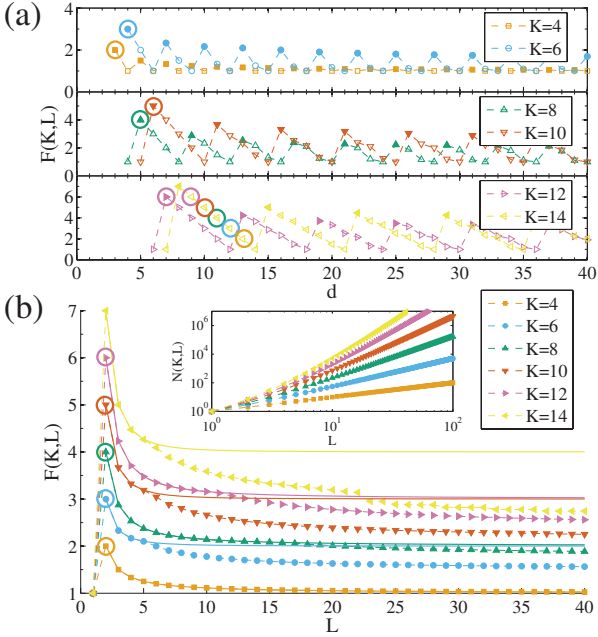}
\caption{\label{fig:mmf_flows}(Colour online). Plot of the sink inflow as a function of (a) the \textit{s-t} distance $d$ and (b) the \textit{s-t} shortest path length $L$ for even $K$ and $4\leq K\leq14$. The relation between $d$ and $L$ can be derived from Eqs.~(\ref{eq:d_star_K,d}) and (\ref{eq:d_star}).
The solid symbols denote the sink inflow values common to both panels. The solid curves in (b) are computed assuming that all bottlenecks are edges of the source $s$ or the sink $t$. The inset in panel (b) shows the approximate polynomial growth of the number of \textit{s-t} shortest paths as $L$ is varied for several $K$ values.
The sink inflow on a $s-t$ pair at the distance $d=d^{min}(K=14,L=2)+i$ for $i=1,\cdots,5$ is the same as the sink inflow on a $s-t$ pair at a smaller distance on a less dense network. The two cases are identified by circles of the same colour on panels (a) and (b).}
\end{figure}

\begin{figure}[phtb]
\includegraphics[width=0.48\textwidth]{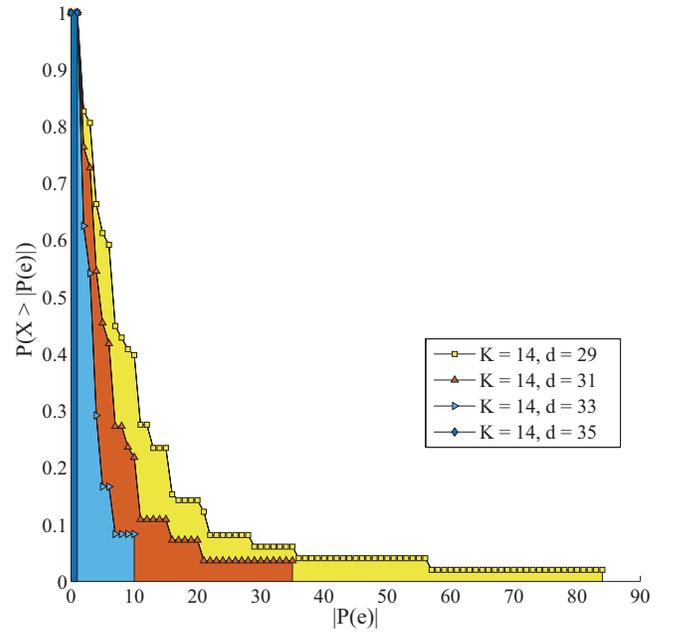}
\caption{\label{fig:mmf_dist}(Colour online). Complementary cumulative distribution of the number of \textit{s-t} shortest paths passing through each edge for $L=5$, as $d$ is varied. We considered only edges that are crossed by at least one path. The diversity of the number of paths crossing each edge is illustrated by the different distributions. When $d=d^{min}(K,L)$, the distribution is broad and some edges are crossed by a large number of paths. The pattern of intersections among these paths constrains the solution of the MMF flow, because the paths share the capacity of the edges they cross. However, when $d= d^{min}(K,L) + K/2 -1$, there is only one \textit{s-t} shortest path. In this case, the MMF algorithm allocates the edge capacity to the path flow, because that path does not interact with any other.}
\end{figure}

Taken together, the methods developed in Sec.~\ref{sec:section_III} can now direct our investigations into 
Max-Min Fair (MMF) flows on nearest neighbour networks with one source and sink pair $(s,t)$. We start by asking how sink inflow varies with node degree $K$, and the distance between the $s$ and $t$ nodes. As discussed in Sec.~\ref{sec:section_III}, we can measure the \textit{s-t} distance in two different ways. On one hand, $L$ is the shortest path length between $s$ and $t$. On the other hand, $d(K,L)$ is the difference between node index labels for $t$ and $s$, as illustrated in Fig.~\ref{fig:visualization_all_panels}, when these labels are increasing from the source to the sink. The two distances are closely related, because  there are $K/2$ \textit{s-t} pairs at distances $d=d^{\min}(K,L),\cdots,d^{\min}(K,L)+K/2-1$, but all of these \textit{s-t} pairs are at the shortest path length $L$ from the source $s$. 
Our first main result from Sec.~\ref{sec:shortest_paths_WS} is that we do not need to investigate the MMF algorithm for all \textit{s-t} distances. In fact, when the \textit{s-t} pair is at a distance $d$ on a $K$-nearest neighbour network, the set of \textit{s-t} shortest paths remains unchanged if the \textit{s-t} pair is placed at a given smaller distance on a sparser network. To be more specific, displacing the \textit{s-t} pair from the distance $d^{\min}(K,L)$ to the distance $d=d^{\min}(K,L)+i$, for $i=1,\cdots,K/2-1$, is equivalent to placing the two nodes at the new distance $d^{\min}(K^\prime,L)$, on a $K^\prime$-nearest neighbour network where  $K^\prime/2 =K/2-i$.
In other words, the \textit{s-t} shortest paths are equivalent for the values of $d$ on the $(K,L)$ network and the values of $K^\prime$ on the sparser $(K^\prime,L)$ network, such that $d+K^\prime/2$ is a constant for a given $L$.
This result implies that the MMF path flows on $K$-nearest neighbour networks are uniquely determined by their values at the distances $d^{\min}(K,L)$, as $K$ and $L$ are varied. This effect can be observed in Fig.~\ref{fig:mmf_flows}a, where we plot the sink inflow as a function of the distance $d$ between the source $s$ and the sink $t$ for several values of $K$. Such transformation from $d^{\min}(K,L)+i$ to $d^{\min}(K^\prime,L)$ explains the oscillatory pattern in the data points. To illustrate this oscillatory effect, we highlight with circles in Fig.~\ref{fig:mmf_flows}a  the sink inflow values for $d=d^{\min}(K=14,L=2)+i$, $i=1,\cdots,5$, as well as the flow values for the corresponding $d^{\min}(K^\prime,L=2)$, demonstrating that the sink inflow is fully determined by its values at the distances $d^{\min}(K,L)$. 

Figure~\ref{fig:mmf_flows}b is a plot of the sink inflow at distances $d^{\min}(K,L)$ as $L$ is varied for several $K$ values, where points common to Figs.~\ref{fig:mmf_flows}a and \ref{fig:mmf_flows}b are represented
by solid symbols. When all bottlenecks are edges of the source or the sink, the sink inflow can be computed with the procedure described in Sec.~\ref{sec:paths_counting}, and the result of such computation is given by the solid curves in Fig.\ref{fig:mmf_flows}b. Indeed, the computation of the path flow increments in Eq.~(\ref{eq:path_flow_at_step_j}), and thus the computation of the sink inflow, can  be done by counting paths that pass through the bottleneck edges, if the location of these bottlenecks is known, and this is our second main result (see Eqs.~(\ref{eq:delta_paths}) and (\ref{eq:sink_inflow2})). 
Furthermore, the path flow increments Eq.~(\ref{eq:delta_paths}) can be determined in closed-form for each $K$ on the region of the $(K,L)$ parameter space where the bottlenecks are edges of $s$ or $t$.

Our choice of shortest path routing allows us to tune the level of interaction among the paths by varying the \textit{s-t} distance $d$ from $d=d^{\min}$ (maximum number of shortest path intersections)  to $d=d^{\min}-K/2+1$ (no intersections). This can be observed in Figs.~\ref{fig:visualization_all_panels} and \ref{fig:sequences}, which illustrate how the number of \textit{s-t} shortest paths depends on $d$. It can also be observed in Fig.~\ref{fig:mmf_dist}, which shows the variation with $d$ of the distribution of the number of \textit{s-t} shortest paths passing through each edge.

Whereas the path flows can be computed in close form if all bottlenecks are edges of $s$ or $t$, it is still possible to find the path flows from path counting methods alone, even when some bottlenecks are in less trivial locations. Indeed, our third main result is a semi-analytical procedure that reduces the computational complexity of the search for bottleneck edges. Brute force approaches to solving numerically the MMF flow assignment for large networks are beyond the capacity of the current generation of computers. Indeed, for $L$ large, the input $N(K,L)$ to the MMF algorithm grows exponentially in size with $K$ (for fixed $L$) and polynomially with $L$ (for fixed $K$), as shown in the inset of Fig.~\ref{fig:mmf_flows}b. The search for bottleneck edges on a generic network is done, at each iteration of the MMF algorithm, over all links of all nodes in the residual network, and that is a procedure of the order $O(\left\vert V\right\vert K)$. Our result simplifies the search procedure for nearest neighbour networks, so that its complexity becomes $O(\left\vert V\right\vert)$, and therefore is independent of node degree $K$. The simplified search procedure consists in finding the minimum in  Eq.~(\ref{eq:path_flow_at_step_j}) over the edges of the residual network that connect two nodes at distance $d=K/2$ (we say that these edges are horizontal). Once the location of  horizontal bottleneck edges is known, the path counting methods developed in Sec.~\ref{sec:section_III} are the key to finding the path flows and the sink inflow. We find the location of the bottleneck edges, as well as the saturated paths, the path flows and the sink inflow, by repeating this procedure for each iteration of the MMF algorithm. 
Nevertheless, the bottlenecks found are not necessarily edges of $s$ or $t$, and we refer to such edges that do not link the source or sink nodes as \textit{inner bottleneck} edges. 
The sink inflow data points in Fig.~\ref{fig:mmf_flows}b were computed with this procedure (also discussed in Sec.~\ref{sec:paths_counting}), and are thus exact results. The inset in Fig.~\ref{fig:mmf_flows}b helps us to appreciate the power of the methodology: our method deals well with  over $10^6$ \textit{s-t} shortest paths for $K=14$ and $L=40$ easily, whereas a brute force version of the MMF algorithm as discussed in Sec.~\ref{sec:paths_counting} would make heavy use of computational resources because all paths and path flows must be stored during the execution of the algorithm. 

\begin{figure}[phtb]
\includegraphics[width=0.48\textwidth]{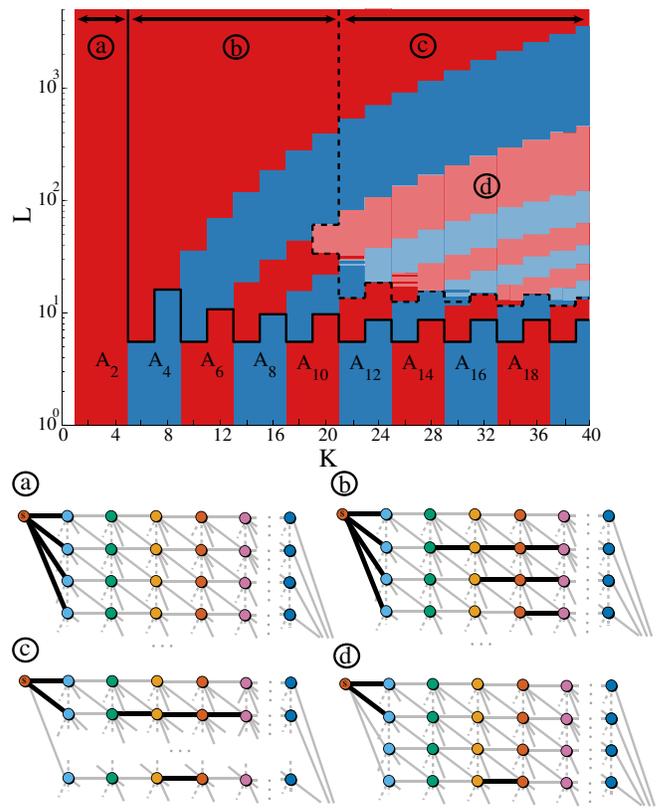}
\caption{\label{fig:diagram}(Colour online). Each cell corresponds to a $(K,L)$ point in the parameter space, where $K$ is even, and the source and sink pair are at a distance $d^{min}(K,L)$ as given by Eq.~(\ref{eq:d_star}). We partition the parameter space $(K,L)$ in areas $A_{(2i)}$, such that after the first $i$ iterations the MMF algorithm finds $2i$ bottlenecks that are edges of the source $s$ or sink $t$ for all cells inside an $A_{(2i)}$ area. We use alternating red and blue coloured cells to distinguish neighbouring $A_{(2i)}$ areas. The parameter space is partitioned into four regions. In region \mycirc{a}, delimited by the solid line, all bottlenecks are edges of the source $s$ or sink $t$. Region \mycirc{b}, between the solid and dashed lines, is characterised by chains of inner bottlenecks followed by a bottleneck at $s$ (or $t$) which still allows us to derive exact results. Region \mycirc{c}, in lighter colours, is defined by the presence of gaps in the row number of consecutive bottleneck edges. Region \mycirc{d} is a subset of \mycirc{c} where the gap occurs between a bottleneck of $s$ or $t$ and the first inner bottleneck.}
\end{figure}

The case for our procedure  becomes even more compelling when we investigate the structure of the parameter space $(K,L)$.
To show this, we construct a parameter space diagram, where we associate with each $(K,L)$ pair the number of bottlenecks that are edges of $s$ or $t$ up to the iteration when the first inner bottleneck is found. The result, which we plot in Fig.~\ref{fig:diagram}, is a partition of the parameter space in areas $A_{2i}$, such that for $K$ and $L$ values in $A_{2i}$ there are $2i$ bottlenecks which are edges of $s$ or $t$.
For $L$ sufficiently large, the behaviour of the MMF algorithm is described by the area $A_2$. In other words, the bottlenecks are edges of $s$ or $t$ only for the first iteration of the MMF algorithm. This means that the two bottlenecks found at the first iteration are edges of $s$ or $t$, but the bottlenecks found at the second iteration are already inner bottlenecks. In fact, when $L$ is large, we only understand well the first iteration of the MMF algorithm, and this opens the possibility that the pattern of the location of bottlenecks in the area $A_2$ of the parameter space may be much more complicated than we have been able to describe. Four regions of the parameter space illustrate what we know about the MMF algorithm in nearest neighbour networks. The first region \mycirc{a}, delimited in Fig.~\ref{fig:diagram} by a black line, is the set of points in the parameter space $(K,L)$ such that all bottlenecks are edges of $s$ or $t$, and this is where we are able to derive analytical results. The second region \mycirc{b}, located between the solid and dashed lines, limits the $(K,L)$ values where the semi-analytical methods developed in Sec.~\ref{sec:paths_counting} can solve all iterations of the MMF algorithm. These semi-analytical methods enable the analysis of the path flows and the sink inflow as $L$ grows. Our main finding is that the calculation of sink inflow for $L$ large in region \mycirc{a} (see Eq.~(\ref{eq:sink_inflow_limit})) is an upper bound to the sink inflow in region \mycirc{b}, as can be observed in Fig.~\ref{fig:mmf_flows}b. This can be explained by the pattern of the location of bottleneck edges in \mycirc{b} (see Fig.~\ref{fig:diagram}). More precisely, the capacity constraints are larger in \mycirc{b} than in \mycirc{a}, because the pattern of bottlenecks in \mycirc{b} contains the pattern of bottlenecks in \mycirc{a}.
The third region \mycirc{c} is characterised by the presence of a gap in the row number of two consecutive bottlenecks on the rectangular subgraph. Region \mycirc{d} is a subset of region \mycirc{c} where the row gap occurs between a bottleneck adjacent to $s$ or $t$. This `gap' in the row number of bottleneck edges implies that a subsequent bottleneck has to appear in the rows that have been skipped. Such detailed knowledge of the location of bottleneck edges in the parameter space of nearest neighbour networks is one of the main contributions of this paper, but it also highlights the limits to our current knowledge of the MMF algorithm, as our methodology is not able to continue computing path flows efficiently for iterations where row gaps occur. The main limitation is  that the implementation of the path counting method  becomes extensively complicated with the appearance of row gaps.

\begin{figure}[phtb]
\includegraphics[width=0.48\textwidth]{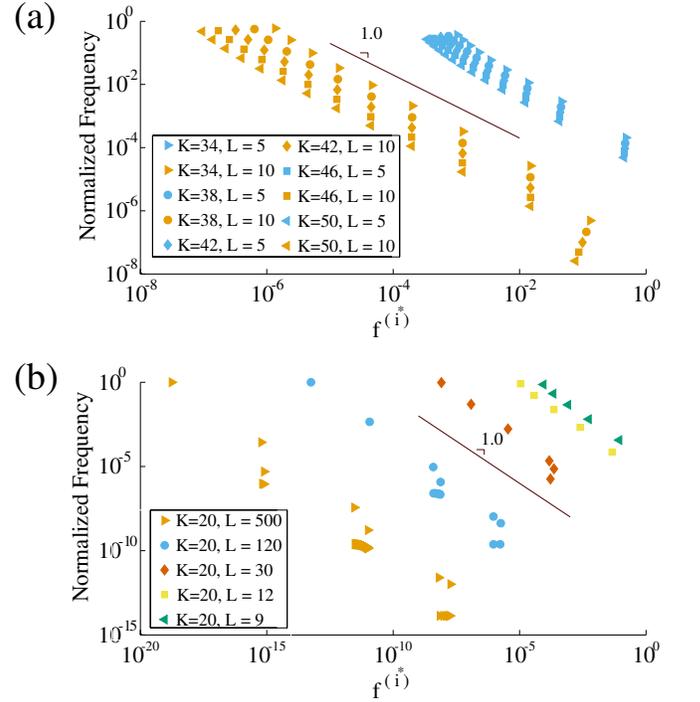}
\caption{\label{fig:histograms}(Colour online). Histograms of the path flow after the MMF algorithm terminates. (a) When the bottlenecks are edges of either the source or sink nodes (or we observe only a small number of inner bottlenecks) the histogram decays approximately as a power law. (b) As the number of inner bottlenecks grows, we observe a  scattered distribution that is caused by different path flows, specific to each inner bottleneck. Two bottlenecks that are on the same row of the rectangular subgraph have similar  path flows. As a consequence, path flows of bottlenecks on the same row are clustered on the histogram.}
\end{figure}

A natural question to ask is then: can network topology influence the Max-Min Fair allocation of path flows? To gain insight into this problem, we plot  the histograms of path flows in Fig.~\ref{fig:histograms} for several $(K,L)$ points inside the regions of the parameter space where we can solve the MMF algorithm exactly (regions \mycirc{a} and \mycirc{b} in Fig.~\ref{fig:histograms}). In the region \mycirc{a} of parameter space where all bottlenecks are edges of $s$ or $t$, the histograms show that the frequency of path flows decays approximately as a power-law with exponent $-1$ (see Fig.~\ref{fig:histograms}a).
Two mechanisms contribute to the way that fair flows are allocated in this area. First, the path flows are dominated by the path flow increment at the last iteration, when the paths are saturated. From Eq.~(\ref{eq:sink_inflow2}), we can write
\begin{equation}
\label{eq:first_factor}
f^{(i^{\ast})}\simeq \Delta f^{(i^{\ast})}= 1/N(K/2-2(i^{\ast}-1),L).
\end{equation}
Second, the number $\eta(i)$ of \textit{s-t} shortest paths that are saturated at each iteration is of the order of magnitude of the number of all paths in the residual network, that is, 
\begin{equation}
\label{eq:second_factor}
\eta^{(i^{\ast})}\simeq N(K/2-2(i^{\ast}-1),L).
\end{equation}
Hence, at each iteration $i^{\ast}$, the path flows, given by Eq.~(\ref{eq:first_factor}), are approximately inversely proportional to the number of saturated paths at iteration $i^{\ast}$, determined by Eq.~(\ref{eq:second_factor}). In other words, $f^{(i^{\ast})}\simeq 1/\eta^{(i^{\ast})}$, and the histogram of path flows decays as a power-law with exponent $-1$, as seen in Fig.~\ref{fig:histograms}a. On the other hand, the presence of inner bottlenecks in the region \mycirc{b} of the parameter space introduces a deviation from the power-law decay (see Fig.~\ref{fig:histograms}b). Indeed, paths crossing the bottlenecks that share the same row of the rectangular subgraph have similar path flows, and thus the frequencies of paths flows are clustered on the histogram.
Taken together, these findings show that power-law allocations can be fair, even when network capacity is uniform, which is a counter-intuitive and unexpected result.

\section{Conclusions}
\label{sec:conclusions}

We analysed the Max-Min Fair flow allocation algorithm on nearest neighbour networks with uniform capacity and node degree $K$. Our study focused on one single source and sink pair, placed at shortest path distance $L$. Transport between the source and the sink can only occur along the shortest paths, each of which is assigned a path flow by the Max-Min Fair algorithm. We note that we have analysed the $(K,L)$ parameter space under the assumption that, for each $L$ value, the \textit{s-t	} pair is at the distance $d=d^{\min} (K,L)$ as given by Eq.~(\ref{eq:d_star}). Other distances $d=d^{\min} (K,L)+i$ (for $1\leq i\leq K/2-1$) can be mapped to an \textit{s-t} pair located at a shorter distance on a sparser network (as discussed in Sec.~\ref{sec:shortest_paths_WS}).

Such configuration led us to analyse how the tuning of the parameters, $K$ and $L$, affects the network throughput between the source and sink pair, as well as the fair allocation of individual path flows. Figure~\ref{fig:mmf_flows} shows the behaviour of the network throughput as a function of the distance between the source and the sink.

We found that the parameter space $(K,L)$ can be partitioned into areas that correspond to distinct patterns in the location of congested edges, \ie~bottlenecks. When the bottlenecks are edges of the source $s$ or sink $t$, we derived a recursive relation that yields the sink inflow and the path flows analytically. Furthermore, in the limit $L\rightarrow\infty$, the sink inflow grows with node degree as the piecewise constant function $\lim_{L\rightarrow\infty}F(K,L)=\lceil K/4\rceil c$. This result is exact for node degree $K\leq 4$, when all bottlenecks are edges of the source or the sink, and we showed that $\lceil K/4\rceil c$ is an upper bound to the sink inflow when $K>4$ and $L$ is large. In contrast, the maximum flow that can be allocated between the source and the sink node pair is $Kc/2$~\cite{Carmi07}. Hence, in the limit $L\rightarrow\infty$ there is a decrease in throughput larger than $50\%$ of max-flow, and this limitation of the Max-Min Fair algorithm is well-known in the literature (see \eg~\cite{Kleinberg01,Bertsimas11}). Indeed, Betsimas \etal~ \cite{Bertsimas11} established an upper bound for the loss of throughput in a very general setup, and when applied to our case with $L\rightarrow\infty$ these results imply $100\%$ reduction of the throughput. Moreover, numerical studies show that when the number of s-t pairs is large and all paths are allowed, the result is a reduction from max-flow comparable to the case of fixed shortest paths~\cite{Nace06}. However, it should be noted that both our results and those of Nace \etal ~\cite{Nace06} are found when paths have many intersections and, hence, are not valid if the network operator would choose the paths not to intersect.
Our results can thus be seen as going further, by providing an exact description of the flow reduction in specific areas of the parameter space. Taken together, these findings suggest that implementations of the Max-Min Fair algorithm in very large real world networks should be restricted to situations where throughput can be sacrificed to achieve fair allocations.

For small values of $L$, all the bottlenecks are edges of $s$ or $t$ and path flows can be computed analytically. As the distance $L$ between the source and the sink increases for a fixed $K$, the pattern of intersections among shortest paths changes and the location of the bottlenecks becomes less regular. To address the new patterns of the location of the bottlenecks, we then derived a semi-analytical method that combines analytical results that reduce the complexity of the search for bottlenecks, with numerical computations for the search. As a result, we were able to solve the Max-Min Fair allocation for large nearest neighbour networks when $K\leq 20$, despite the very large number of shortest paths involved in the computations.

When all bottlenecks are edges of the source or the sink, we find that the histogram of path flows decays approximately as a power-law with exponent $-1$. This uneven distribution is unforeseen, because the Max-Min Fair algorithm assigns path flows fairly, \ie~the least well off get as much as possible under the capacity constraints. This counter-intuitive result that power-law allocations can be fair is a consequence of the constraints placed on the Max-Min Fair allocation by the combination of the network topology and the routing over shortest paths.
Moreover, we observed deviations from this power-law decay in regions of the parameter space where some bottlenecks are not edges of the source or sink.
These unexpected results suggest that network designers should be aware of the interplay between the structure of transport routes and network topology, which is a crucial factor in the fair allocation of network flows.

Finally, the analysis developed in this paper may be extended 
in several directions. A possible line of enquiry  
is to study the network throughput as a function of the number of distinct sink-source 
pairs. Still another possible step is to investigate the effect of network topology on Max-Min Fair flows. Further studies are required to understand whether the reduction in throughput that we have found is specific to the choice of fixed shortest paths, or whether it is a more general result. Our study is just a first step towards a rigorous understanding of fair allocations in complex networks. Due to the importance of the topic, further insights could be reached by analysing random graphs or real world network topologies. It would also be important to gain a better understanding of the relation between different routing strategies and the network throughput. Last, but not least, it would be worthwhile to consider other fair allocation schemes, such as the Proportional Fairness~\cite{Kelly98}.

\begin{acknowledgments}
We gratefully acknowledge the support of the RAVEN EPSRC project (EP/H04812X/1).  L.B.  gratefully acknowledges partial financial support by the Ministry of Education of the Slovak Republic (project VEGA 1/0296/12).
\end{acknowledgments}


\bibliographystyle{apsrev}
\bibliography{arxiv}

\begin{thebibliography}{49}
\expandafter\ifx\csname natexlab\endcsname\relax\def\natexlab#1{#1}\fi
\expandafter\ifx\csname bibnamefont\endcsname\relax
  \def\bibnamefont#1{#1}\fi
\expandafter\ifx\csname bibfnamefont\endcsname\relax
  \def\bibfnamefont#1{#1}\fi
\expandafter\ifx\csname citenamefont\endcsname\relax
  \def\citenamefont#1{#1}\fi
\expandafter\ifx\csname url\endcsname\relax
  \def\url#1{\texttt{#1}}\fi
\expandafter\ifx\csname urlprefix\endcsname\relax\def\urlprefix{URL }\fi
\providecommand{\bibinfo}[2]{#2}
\providecommand{\eprint}[2][]{\url{#2}}

\bibitem[{\citenamefont{Whittle}(2007)}]{Whittle07}
\bibinfo{author}{\bibfnamefont{P.}~\bibnamefont{Whittle}},
  \emph{\bibinfo{title}{Networks Optimisation and Evolution}}, Cambridge Series
  in Statistical and Probabilistic Mathematics (\bibinfo{publisher}{Cambridge
  University Press}, \bibinfo{year}{2007}).

\bibitem[{\citenamefont{Barrat et~al.}(2008)\citenamefont{Barrat,
  Barth\'{e}lemy, and Vespignani}}]{Barrat08}
\bibinfo{author}{\bibfnamefont{A.}~\bibnamefont{Barrat}},
  \bibinfo{author}{\bibfnamefont{M.}~\bibnamefont{Barth\'{e}lemy}},
  \bibnamefont{and}
  \bibinfo{author}{\bibfnamefont{A.}~\bibnamefont{Vespignani}},
  \emph{\bibinfo{title}{Dynamical Processes on Complex Networks}}
  (\bibinfo{publisher}{Cambridge University Press}, \bibinfo{year}{2008}).

\bibitem[{\citenamefont{Bohn and Magnasco}(2007)}]{Bohn07}
\bibinfo{author}{\bibfnamefont{S.}~\bibnamefont{Bohn}} \bibnamefont{and}
  \bibinfo{author}{\bibfnamefont{M.~O.} \bibnamefont{Magnasco}},
  \bibinfo{journal}{Phys. Rev. Lett.} \textbf{\bibinfo{volume}{98}},
  \bibinfo{pages}{088702} (\bibinfo{year}{2007}).

\bibitem[{\citenamefont{Corson}(2010)}]{Corson10}
\bibinfo{author}{\bibfnamefont{F.}~\bibnamefont{Corson}},
  \bibinfo{journal}{Phys. Rev. Lett.} \textbf{\bibinfo{volume}{104}},
  \bibinfo{pages}{048703} (\bibinfo{year}{2010}).

\bibitem[{\citenamefont{Li et~al.}(2010)\citenamefont{Li, Reis, Moreira,
  Havlin, Stanley, and Andrade}}]{Li10}
\bibinfo{author}{\bibfnamefont{G.}~\bibnamefont{Li}},
  \bibinfo{author}{\bibfnamefont{S.~D.~S.} \bibnamefont{Reis}},
  \bibinfo{author}{\bibfnamefont{A.~A.} \bibnamefont{Moreira}},
  \bibinfo{author}{\bibfnamefont{S.}~\bibnamefont{Havlin}},
  \bibinfo{author}{\bibfnamefont{H.~E.} \bibnamefont{Stanley}},
  \bibnamefont{and} \bibinfo{author}{\bibfnamefont{J.~S.}
  \bibnamefont{Andrade}}, \bibinfo{journal}{Phys. Rev. Lett.}
  \textbf{\bibinfo{volume}{104}}, \bibinfo{pages}{018701}
  (\bibinfo{year}{2010}).

\bibitem[{\citenamefont{Xia and Hill}(2010)}]{Xia10}
\bibinfo{author}{\bibfnamefont{Y.~X.} \bibnamefont{Xia}} \bibnamefont{and}
  \bibinfo{author}{\bibfnamefont{D.}~\bibnamefont{Hill}},
  \bibinfo{journal}{EPL} \textbf{\bibinfo{volume}{89}}, \bibinfo{pages}{58004}
  (\bibinfo{year}{2010}).

\bibitem[{\citenamefont{Carvalho et~al.}(2009)\citenamefont{Carvalho, Buzna,
  Bono, Gutierrez, Just, and Arrowsmith}}]{Carvalho09}
\bibinfo{author}{\bibfnamefont{R.}~\bibnamefont{Carvalho}},
  \bibinfo{author}{\bibfnamefont{L.}~\bibnamefont{Buzna}},
  \bibinfo{author}{\bibfnamefont{F.}~\bibnamefont{Bono}},
  \bibinfo{author}{\bibfnamefont{E.}~\bibnamefont{Gutierrez}},
  \bibinfo{author}{\bibfnamefont{W.}~\bibnamefont{Just}}, \bibnamefont{and}
  \bibinfo{author}{\bibfnamefont{D.}~\bibnamefont{Arrowsmith}},
  \bibinfo{journal}{Phys. Rev. E} \textbf{\bibinfo{volume}{80}},
  \bibinfo{pages}{016106} (\bibinfo{year}{2009}).

\bibitem[{\citenamefont{Katifori et~al.}(2010)\citenamefont{Katifori, Szollosi,
  and Magnasco}}]{Katifori10}
\bibinfo{author}{\bibfnamefont{E.}~\bibnamefont{Katifori}},
  \bibinfo{author}{\bibfnamefont{G.~J.} \bibnamefont{Szollosi}},
  \bibnamefont{and} \bibinfo{author}{\bibfnamefont{M.~O.}
  \bibnamefont{Magnasco}}, \bibinfo{journal}{Phys. Rev. Lett.}
  \textbf{\bibinfo{volume}{104}}, \bibinfo{pages}{048704}
  (\bibinfo{year}{2010}).

\bibitem[{\citenamefont{Tero et~al.}(2010)\citenamefont{Tero, Takagi, Saigusa,
  Ito, Bebber, Fricker, Yumiki, Kobayashi, and Nakagaki}}]{Tero10}
\bibinfo{author}{\bibfnamefont{A.}~\bibnamefont{Tero}},
  \bibinfo{author}{\bibfnamefont{S.}~\bibnamefont{Takagi}},
  \bibinfo{author}{\bibfnamefont{T.}~\bibnamefont{Saigusa}},
  \bibinfo{author}{\bibfnamefont{K.}~\bibnamefont{Ito}},
  \bibinfo{author}{\bibfnamefont{D.~P.} \bibnamefont{Bebber}},
  \bibinfo{author}{\bibfnamefont{M.~D.} \bibnamefont{Fricker}},
  \bibinfo{author}{\bibfnamefont{K.}~\bibnamefont{Yumiki}},
  \bibinfo{author}{\bibfnamefont{R.}~\bibnamefont{Kobayashi}},
  \bibnamefont{and} \bibinfo{author}{\bibfnamefont{T.}~\bibnamefont{Nakagaki}},
  \bibinfo{journal}{Science} \textbf{\bibinfo{volume}{327}},
  \bibinfo{pages}{439} (\bibinfo{year}{2010}).

\bibitem[{\citenamefont{Ohira and Sawatari}(1998)}]{Ohira98}
\bibinfo{author}{\bibfnamefont{T.}~\bibnamefont{Ohira}} \bibnamefont{and}
  \bibinfo{author}{\bibfnamefont{R.}~\bibnamefont{Sawatari}},
  \bibinfo{journal}{Phys. Rev. E} \textbf{\bibinfo{volume}{58}},
  \bibinfo{pages}{193} (\bibinfo{year}{1998}).

\bibitem[{\citenamefont{Danila et~al.}(2006)\citenamefont{Danila, Yu, Marsh,
  and Bassler}}]{Danila06}
\bibinfo{author}{\bibfnamefont{B.}~\bibnamefont{Danila}},
  \bibinfo{author}{\bibfnamefont{Y.}~\bibnamefont{Yu}},
  \bibinfo{author}{\bibfnamefont{J.~A.} \bibnamefont{Marsh}}, \bibnamefont{and}
  \bibinfo{author}{\bibfnamefont{K.~E.} \bibnamefont{Bassler}},
  \bibinfo{journal}{Phys. Rev. E} \textbf{\bibinfo{volume}{74}},
  \bibinfo{pages}{046106} (\bibinfo{year}{2006}).

\bibitem[{\citenamefont{Sreenivasan et~al.}(2007)\citenamefont{Sreenivasan,
  Cohen, Lopez, Toroczkai, and Stanley}}]{Sreenivasan07}
\bibinfo{author}{\bibfnamefont{S.}~\bibnamefont{Sreenivasan}},
  \bibinfo{author}{\bibfnamefont{R.}~\bibnamefont{Cohen}},
  \bibinfo{author}{\bibfnamefont{E.}~\bibnamefont{Lopez}},
  \bibinfo{author}{\bibfnamefont{Z.}~\bibnamefont{Toroczkai}},
  \bibnamefont{and} \bibinfo{author}{\bibfnamefont{H.~E.}
  \bibnamefont{Stanley}}, \bibinfo{journal}{Phys. Rev. E}
  \textbf{\bibinfo{volume}{75}}, \bibinfo{pages}{036105}
  (\bibinfo{year}{2007}).

\bibitem[{\citenamefont{De~Martino et~al.}(2009)\citenamefont{De~Martino,
  Dall'Asta, Bianconi, and Marsili}}]{Marsili09}
\bibinfo{author}{\bibfnamefont{D.}~\bibnamefont{De~Martino}},
  \bibinfo{author}{\bibfnamefont{L.}~\bibnamefont{Dall'Asta}},
  \bibinfo{author}{\bibfnamefont{G.}~\bibnamefont{Bianconi}}, \bibnamefont{and}
  \bibinfo{author}{\bibfnamefont{M.}~\bibnamefont{Marsili}},
  \bibinfo{journal}{Phys. Rev. E} \textbf{\bibinfo{volume}{79}},
  \bibinfo{pages}{015101(R)} (\bibinfo{year}{2009}).

\bibitem[{\citenamefont{Danila et~al.}(2009)\citenamefont{Danila, Sun, and
  Bassler}}]{Danila09}
\bibinfo{author}{\bibfnamefont{B.}~\bibnamefont{Danila}},
  \bibinfo{author}{\bibfnamefont{Y.~D.} \bibnamefont{Sun}}, \bibnamefont{and}
  \bibinfo{author}{\bibfnamefont{K.~E.} \bibnamefont{Bassler}},
  \bibinfo{journal}{Phys. Rev. E} \textbf{\bibinfo{volume}{80}},
  \bibinfo{pages}{066116} (\bibinfo{year}{2009}).

\bibitem[{\citenamefont{Helbing et~al.}(2002)\citenamefont{Helbing, Schonhof,
  and Kern}}]{Helbing02}
\bibinfo{author}{\bibfnamefont{D.}~\bibnamefont{Helbing}},
  \bibinfo{author}{\bibfnamefont{M.}~\bibnamefont{Schonhof}}, \bibnamefont{and}
  \bibinfo{author}{\bibfnamefont{D.}~\bibnamefont{Kern}}, \bibinfo{journal}{New
  J. Phys.} \textbf{\bibinfo{volume}{4}}, \bibinfo{pages}{33}
  (\bibinfo{year}{2002}).

\bibitem[{\citenamefont{Helbing et~al.}(2005)\citenamefont{Helbing, Schonhof,
  Stark, and Holyst}}]{Helbing05}
\bibinfo{author}{\bibfnamefont{D.}~\bibnamefont{Helbing}},
  \bibinfo{author}{\bibfnamefont{M.}~\bibnamefont{Schonhof}},
  \bibinfo{author}{\bibfnamefont{H.~U.} \bibnamefont{Stark}}, \bibnamefont{and}
  \bibinfo{author}{\bibfnamefont{J.~A.} \bibnamefont{Holyst}},
  \bibinfo{journal}{Adv. Complex Syst.} \textbf{\bibinfo{volume}{8}},
  \bibinfo{pages}{87} (\bibinfo{year}{2005}).

\bibitem[{\citenamefont{Youn et~al.}(2008)\citenamefont{Youn, Gastner, and
  Jeong}}]{Youn08}
\bibinfo{author}{\bibfnamefont{H.}~\bibnamefont{Youn}},
  \bibinfo{author}{\bibfnamefont{M.~T.} \bibnamefont{Gastner}},
  \bibnamefont{and} \bibinfo{author}{\bibfnamefont{H.}~\bibnamefont{Jeong}},
  \bibinfo{journal}{Phys. Rev. Lett.} \textbf{\bibinfo{volume}{101}},
  \bibinfo{pages}{128701} (\bibinfo{year}{2008}).

\bibitem[{\citenamefont{Bertsimas et~al.}(2011)\citenamefont{Bertsimas, Farias,
  and Trichakis}}]{Bertsimas11}
\bibinfo{author}{\bibfnamefont{D.}~\bibnamefont{Bertsimas}},
  \bibinfo{author}{\bibfnamefont{V.~F.} \bibnamefont{Farias}},
  \bibnamefont{and}
  \bibinfo{author}{\bibfnamefont{N.}~\bibnamefont{Trichakis}},
  \bibinfo{journal}{Oper. Res.} \textbf{\bibinfo{volume}{59}},
  \bibinfo{pages}{17} (\bibinfo{year}{2011}).

\bibitem[{\citenamefont{Brams and Taylor}(1996)}]{Brams96}
\bibinfo{author}{\bibfnamefont{S.~J.} \bibnamefont{Brams}} \bibnamefont{and}
  \bibinfo{author}{\bibfnamefont{A.~D.} \bibnamefont{Taylor}},
  \emph{\bibinfo{title}{Fair Division: From cake-cutting to dispute
  resolution}} (\bibinfo{publisher}{Cambridge University Press},
  \bibinfo{year}{1996}).

\bibitem[{\citenamefont{Robertson and Webb}(1998)}]{Robertson98}
\bibinfo{author}{\bibfnamefont{J.}~\bibnamefont{Robertson}} \bibnamefont{and}
  \bibinfo{author}{\bibfnamefont{W.}~\bibnamefont{Webb}},
  \emph{\bibinfo{title}{Cake-Cutting Algorithms: Be Fair If You Can}}
  (\bibinfo{publisher}{A K Peters}, \bibinfo{year}{1998}).

\bibitem[{\citenamefont{Barbanel}(2005)}]{Barbanel05}
\bibinfo{author}{\bibfnamefont{J.}~\bibnamefont{Barbanel}},
  \emph{\bibinfo{title}{The Geometry of Efficient Fair Division}}
  (\bibinfo{publisher}{Cambridge University Press}, \bibinfo{year}{2005}).

\bibitem[{\citenamefont{Taylor and Pacelli}(2008)}]{Taylor08}
\bibinfo{author}{\bibfnamefont{A.~D.} \bibnamefont{Taylor}} \bibnamefont{and}
  \bibinfo{author}{\bibfnamefont{A.~M.} \bibnamefont{Pacelli}},
  \emph{\bibinfo{title}{Mathematics and Politics: Strategy, Voting, Power and
  Proof}} (\bibinfo{publisher}{Springer}, \bibinfo{year}{2008}).

\bibitem[{\citenamefont{Bertsekas and Gallager}(1992)}]{Bertsekas92}
\bibinfo{author}{\bibfnamefont{D.~P.} \bibnamefont{Bertsekas}}
  \bibnamefont{and} \bibinfo{author}{\bibfnamefont{R.}~\bibnamefont{Gallager}},
  \emph{\bibinfo{title}{Data Networks}} (\bibinfo{publisher}{Prentice Hall},
  \bibinfo{year}{1992}), \bibinfo{edition}{2nd} ed.

\bibitem[{\citenamefont{Kelly et~al.}(1998)\citenamefont{Kelly, Maulloo, and
  Tan}}]{Kelly98}
\bibinfo{author}{\bibfnamefont{F.~P.} \bibnamefont{Kelly}},
  \bibinfo{author}{\bibfnamefont{A.~K.} \bibnamefont{Maulloo}},
  \bibnamefont{and} \bibinfo{author}{\bibfnamefont{D.~K.~H.}
  \bibnamefont{Tan}}, \bibinfo{journal}{J. Oper. Res. Soc.}
  \textbf{\bibinfo{volume}{49}}, \bibinfo{pages}{237} (\bibinfo{year}{1998}).

\bibitem[{\citenamefont{Srikant}(2003)}]{Srikant03}
\bibinfo{author}{\bibfnamefont{R.}~\bibnamefont{Srikant}},
  \emph{\bibinfo{title}{The Mathematics of Internet Congestion Control}}
  (\bibinfo{publisher}{Birkh\"{a}user}, \bibinfo{year}{2003}).

\bibitem[{\citenamefont{Kelly}(2000)}]{Kelly00}
\bibinfo{author}{\bibfnamefont{F.~P.} \bibnamefont{Kelly}},
  \bibinfo{journal}{Philos. Trans. R. Soc. Lond. Ser. A-Math. Phys. Eng. Sci.}
  \textbf{\bibinfo{volume}{358}}, \bibinfo{pages}{2335} (\bibinfo{year}{2000}).

\bibitem[{\citenamefont{Kleinberg et~al.}(2001)\citenamefont{Kleinberg, Rabani,
  and Tardos}}]{Kleinberg01}
\bibinfo{author}{\bibfnamefont{J.}~\bibnamefont{Kleinberg}},
  \bibinfo{author}{\bibfnamefont{Y.}~\bibnamefont{Rabani}}, \bibnamefont{and}
  \bibinfo{author}{\bibfnamefont{E.}~\bibnamefont{Tardos}},
  \bibinfo{journal}{J. Comput. Syst. Sci.} \textbf{\bibinfo{volume}{63}},
  \bibinfo{pages}{2} (\bibinfo{year}{2001}).

\bibitem[{\citenamefont{Massoulie and Roberts}(2002)}]{Massoulie02}
\bibinfo{author}{\bibfnamefont{L.}~\bibnamefont{Massoulie}} \bibnamefont{and}
  \bibinfo{author}{\bibfnamefont{J.}~\bibnamefont{Roberts}},
  \bibinfo{journal}{IEEE-ACM Trans. Netw.} \textbf{\bibinfo{volume}{10}},
  \bibinfo{pages}{320} (\bibinfo{year}{2002}).

\bibitem[{\citenamefont{Megiddo}(1974)}]{Megiddo74}
\bibinfo{author}{\bibfnamefont{N.}~\bibnamefont{Megiddo}},
  \bibinfo{journal}{Mathematical Programming} \textbf{\bibinfo{volume}{7}},
  \bibinfo{pages}{97} (\bibinfo{year}{1974}).

\bibitem[{\citenamefont{Megiddo}(1977)}]{Megiddo77}
\bibinfo{author}{\bibfnamefont{N.}~\bibnamefont{Megiddo}},
  \bibinfo{journal}{Bull. Amer. Math. Soc.} \textbf{\bibinfo{volume}{83}},
  \bibinfo{pages}{407} (\bibinfo{year}{1977}).

\bibitem[{\citenamefont{Wong and Saad}(2007)}]{Saad07}
\bibinfo{author}{\bibfnamefont{K.~Y.~M.} \bibnamefont{Wong}} \bibnamefont{and}
  \bibinfo{author}{\bibfnamefont{D.}~\bibnamefont{Saad}},
  \bibinfo{journal}{Phys. Rev. E} \textbf{\bibinfo{volume}{76}},
  \bibinfo{pages}{011115} (\bibinfo{year}{2007}).

\bibitem[{\citenamefont{Jaffe}(1981)}]{Jaffe81}
\bibinfo{author}{\bibfnamefont{J.~M.} \bibnamefont{Jaffe}},
  \bibinfo{journal}{IEEE Trans. Commun.} \textbf{\bibinfo{volume}{29}},
  \bibinfo{pages}{954} (\bibinfo{year}{1981}).

\bibitem[{\citenamefont{Kumar and Kleinberg}(2006)}]{Kumar06}
\bibinfo{author}{\bibfnamefont{A.}~\bibnamefont{Kumar}} \bibnamefont{and}
  \bibinfo{author}{\bibfnamefont{J.}~\bibnamefont{Kleinberg}},
  \bibinfo{journal}{SIAM J. Comput.} \textbf{\bibinfo{volume}{36}},
  \bibinfo{pages}{657} (\bibinfo{year}{2006}).

\bibitem[{\citenamefont{Anshelevich et~al.}(2008)\citenamefont{Anshelevich,
  Dasgupta, Kleinberg, Tardos, Wexler, and Roughgarden}}]{Anshelevich08}
\bibinfo{author}{\bibfnamefont{E.}~\bibnamefont{Anshelevich}},
  \bibinfo{author}{\bibfnamefont{A.}~\bibnamefont{Dasgupta}},
  \bibinfo{author}{\bibfnamefont{J.}~\bibnamefont{Kleinberg}},
  \bibinfo{author}{\bibfnamefont{E.}~\bibnamefont{Tardos}},
  \bibinfo{author}{\bibfnamefont{T.}~\bibnamefont{Wexler}}, \bibnamefont{and}
  \bibinfo{author}{\bibfnamefont{T.}~\bibnamefont{Roughgarden}},
  \bibinfo{journal}{SIAM J. Comput.} \textbf{\bibinfo{volume}{38}},
  \bibinfo{pages}{1602} (\bibinfo{year}{2008}).

\bibitem[{\citenamefont{Nace et~al.}(2006)\citenamefont{Nace, Doan, Gourdin,
  and Liau}}]{Nace06}
\bibinfo{author}{\bibfnamefont{D.}~\bibnamefont{Nace}},
  \bibinfo{author}{\bibfnamefont{N.~L.} \bibnamefont{Doan}},
  \bibinfo{author}{\bibfnamefont{E.}~\bibnamefont{Gourdin}}, \bibnamefont{and}
  \bibinfo{author}{\bibfnamefont{B.}~\bibnamefont{Liau}},
  \bibinfo{journal}{IEEE-ACM Trans. Netw.} \textbf{\bibinfo{volume}{14}},
  \bibinfo{pages}{1272} (\bibinfo{year}{2006}).

\bibitem[{\citenamefont{Nace et~al.}(2008)\citenamefont{Nace, Doan,
  Klopfenstein, and Bashllari}}]{Nace08a}
\bibinfo{author}{\bibfnamefont{D.}~\bibnamefont{Nace}},
  \bibinfo{author}{\bibfnamefont{L.~N.} \bibnamefont{Doan}},
  \bibinfo{author}{\bibfnamefont{O.}~\bibnamefont{Klopfenstein}},
  \bibnamefont{and}
  \bibinfo{author}{\bibfnamefont{A.}~\bibnamefont{Bashllari}},
  \bibinfo{journal}{Comput. Oper. Res.} \textbf{\bibinfo{volume}{35}},
  \bibinfo{pages}{557} (\bibinfo{year}{2008}).

\bibitem[{\citenamefont{Radunovic and Le~Boudec}(2007)}]{Radunovic07}
\bibinfo{author}{\bibfnamefont{B.}~\bibnamefont{Radunovic}} \bibnamefont{and}
  \bibinfo{author}{\bibfnamefont{J.~Y.} \bibnamefont{Le~Boudec}},
  \bibinfo{journal}{IEEE-ACM Trans. Netw.} \textbf{\bibinfo{volume}{15}},
  \bibinfo{pages}{1073} (\bibinfo{year}{2007}).

\bibitem[{\citenamefont{Nace and Pioro}(2008)}]{Nace08}
\bibinfo{author}{\bibfnamefont{D.}~\bibnamefont{Nace}} \bibnamefont{and}
  \bibinfo{author}{\bibfnamefont{M.}~\bibnamefont{Pioro}},
  \bibinfo{journal}{IEEE Commun. Surv. Tutor.} \textbf{\bibinfo{volume}{10}},
  \bibinfo{pages}{5} (\bibinfo{year}{2008}).

\bibitem[{\citenamefont{Costa et~al.}(2011)\citenamefont{Costa, Oliveira,
  Travieso, Rodrigues, Boas, Antiqueira, Viana, and Rocha}}]{Costa11}
\bibinfo{author}{\bibfnamefont{L.~D.} \bibnamefont{Costa}},
  \bibinfo{author}{\bibfnamefont{O.~N.} \bibnamefont{Oliveira}},
  \bibinfo{author}{\bibfnamefont{G.}~\bibnamefont{Travieso}},
  \bibinfo{author}{\bibfnamefont{F.~A.} \bibnamefont{Rodrigues}},
  \bibinfo{author}{\bibfnamefont{P.~R.~V.} \bibnamefont{Boas}},
  \bibinfo{author}{\bibfnamefont{L.}~\bibnamefont{Antiqueira}},
  \bibinfo{author}{\bibfnamefont{M.~P.} \bibnamefont{Viana}}, \bibnamefont{and}
  \bibinfo{author}{\bibfnamefont{L.~E.~C.} \bibnamefont{Rocha}},
  \bibinfo{journal}{Adv. Phys.} \textbf{\bibinfo{volume}{60}},
  \bibinfo{pages}{329} (\bibinfo{year}{2011}).

\bibitem[{\citenamefont{Boccaletti et~al.}(2006)\citenamefont{Boccaletti,
  Latora, Moreno, Chavez, and Hwang}}]{Boccaletti_PR_06}
\bibinfo{author}{\bibfnamefont{S.}~\bibnamefont{Boccaletti}},
  \bibinfo{author}{\bibfnamefont{V.}~\bibnamefont{Latora}},
  \bibinfo{author}{\bibfnamefont{Y.}~\bibnamefont{Moreno}},
  \bibinfo{author}{\bibfnamefont{M.}~\bibnamefont{Chavez}}, \bibnamefont{and}
  \bibinfo{author}{\bibfnamefont{D.~U.} \bibnamefont{Hwang}},
  \bibinfo{journal}{Phys. Rep.} \textbf{\bibinfo{volume}{424}},
  \bibinfo{pages}{175} (\bibinfo{year}{2006}).

\bibitem[{\citenamefont{de~Menezes and Barabasi}(2004)}]{Menezes04}
\bibinfo{author}{\bibfnamefont{M.~A.} \bibnamefont{de~Menezes}}
  \bibnamefont{and} \bibinfo{author}{\bibfnamefont{A.~L.}
  \bibnamefont{Barabasi}}, \bibinfo{journal}{Phys. Rev. Lett.}
  \textbf{\bibinfo{volume}{92}}, \bibinfo{pages}{028701}
  (\bibinfo{year}{2004}).

\bibitem[{\citenamefont{Gastner and Newman}(2006)}]{Gastner06a}
\bibinfo{author}{\bibfnamefont{M.~T.} \bibnamefont{Gastner}} \bibnamefont{and}
  \bibinfo{author}{\bibfnamefont{M.~E.~J.} \bibnamefont{Newman}},
  \bibinfo{journal}{J. Stat. Mech.} p. \bibinfo{pages}{P01015}
  (\bibinfo{year}{2006}).

\bibitem[{\citenamefont{Latora and Marchiori}(2001)}]{Latora01}
\bibinfo{author}{\bibfnamefont{V.}~\bibnamefont{Latora}} \bibnamefont{and}
  \bibinfo{author}{\bibfnamefont{M.}~\bibnamefont{Marchiori}},
  \bibinfo{journal}{Phys. Rev. Lett.} \textbf{\bibinfo{volume}{87}},
  \bibinfo{pages}{198701} (\bibinfo{year}{2001}).

\bibitem[{\citenamefont{Pioro and Medhi}(2004)}]{Pioro04}
\bibinfo{author}{\bibfnamefont{M.}~\bibnamefont{Pioro}} \bibnamefont{and}
  \bibinfo{author}{\bibfnamefont{D.}~\bibnamefont{Medhi}},
  \emph{\bibinfo{title}{Routing, Flow, and Capacity Design in Communication and
  Computer Networks}} (\bibinfo{publisher}{Morgan Kaufmann},
  \bibinfo{year}{2004}).

\bibitem[{Note1()}]{Note1}
\bibinfo{note}{An alternative would be to consider a directed network,
  and path flows as a snapshot of network usage in time, but we will not
  explore this possibility here because undirected networks are traditionally
  studied before directed ones.}

\bibitem[{\citenamefont{Watts and Strogatz}(1998)}]{Watts98}
\bibinfo{author}{\bibfnamefont{D.~J.} \bibnamefont{Watts}} \bibnamefont{and}
  \bibinfo{author}{\bibfnamefont{S.~H.} \bibnamefont{Strogatz}},
  \bibinfo{journal}{Nature} \textbf{\bibinfo{volume}{393}},
  \bibinfo{pages}{440} (\bibinfo{year}{1998}).

\bibitem[{\citenamefont{Newman}(2010)}]{Newman2010}
\bibinfo{author}{\bibfnamefont{M.}~\bibnamefont{Newman}},
  \emph{\bibinfo{title}{Networks: An Introduction}} (\bibinfo{publisher}{Oxford
  University Press}, \bibinfo{year}{2010}).

\bibitem[{\citenamefont{Stanley}(2000)}]{RichardStanley00}
\bibinfo{author}{\bibfnamefont{R.~P.} \bibnamefont{Stanley}},
  \emph{\bibinfo{title}{Enumerative Combinatorics}}, vol.~\bibinfo{volume}{1}
  (\bibinfo{publisher}{Cambridge University Press}, \bibinfo{year}{2000}),
  \bibinfo{edition}{2nd} ed.

\bibitem[{\citenamefont{Carmi et~al.}(2007)\citenamefont{Carmi, Wu, Lopez,
  Havlin, and Stanley}}]{Carmi07}
\bibinfo{author}{\bibfnamefont{S.}~\bibnamefont{Carmi}},
  \bibinfo{author}{\bibfnamefont{Z.}~\bibnamefont{Wu}},
  \bibinfo{author}{\bibfnamefont{E.}~\bibnamefont{Lopez}},
  \bibinfo{author}{\bibfnamefont{S.}~\bibnamefont{Havlin}}, \bibnamefont{and}
  \bibinfo{author}{\bibfnamefont{H.~E.} \bibnamefont{Stanley}},
  \bibinfo{journal}{Eur. Phys. J. B} \textbf{\bibinfo{volume}{57}},
  \bibinfo{pages}{165} (\bibinfo{year}{2007}).

\end{thebibliography}

\end{document}